\documentclass[structabstract]{aa} 
\usepackage{multirow}
\usepackage{natbib,twoopt}
\usepackage{natbib}
\usepackage{fixltx2e}
\bibpunct{(}{)}{;}{a}{}{,} 
\usepackage{graphicx}
\usepackage{txfonts}

\begin{document}

\title{Non-Zeeman Circular Polarization of CO rotational lines in \\ SNR IC 443}
\titlerunning{Non-Zeeman Circular Polarization of spectral lines}

\author{Talayeh Hezareh \inst{1}
             \and Helmut Wiesemeyer \inst{1} 
             \and Martin Houde \inst{2,3} 
             \and Antoine Gusdorf \inst{4} 
             \and Giorgio Siringo \inst{5}}

\institute{Max-Planck-Institut f\"ur Radioastronomie, Auf dem
  H\"{u}gel 69, 53121 Bonn, Germany \\
           \email{thezareh@mpifr-bonn.mpg.de}\label{inst1} \and
           Department of Physics and Astronomy, The University of
           Western Ontario, London, Ontario, Canada, N6A
           3K7 \label{inst2} \and  
					 Division of Physics,
                                         Mathematics and Astronomy,
                                         California Institute of
                                         Technology, Pasadena, CA
                                         91125, U.S.A. 
					 \label{inst3} \and
                                         LERMA, UMR 8112 du CNRS,
                                         Observatoire de Paris,
                                         \'Ecole Normale Sup\'erieure,
                                         24 rue Lhomond, F75231 Paris
                                         Cedex 05, France  \label{inst4}  \and 
                                         European Southern Observatory
                                         and Joint ALMA Observatory,
                                         Chile  \label{inst5}}


\abstract
{We investigate non-Zeeman circular
  polarization and linear polarization levels of up to $1\%$ of $^{12}$CO spectral line emission
  detected in a shocked molecular clump
  around the supernova remnant (SNR) IC 443, with the goal of understanding the magnetic field
  structure in this source.}  
{We examine our polarization results to confirm that the circular polarization
  signal in CO lines is caused by a conversion of linear to circular
  polarization, consistent with anisotropic resonant scattering. In
  this process background linearly
  polarized CO emission interacts with similar foreground molecules aligned with the
  ambient magnetic field and scatters at a transition frequency. The
  difference in phase shift  between the orthogonally polarized components of 
  this scattered emission can cause a transformation of linear to
  circular polarization.} 
 {We compared linear polarization maps from 
  dust continuum, obtained with PolKa at APEX, and $^{12}$CO ($J=2\rightarrow1$) and
  ($J=1\rightarrow0$) from the IRAM 30-m telescope and found
  no consistency between the two sets of polarization maps. We then  
  reinserted the measured circular polarization signal in the CO lines
  across the source to the corresponding linear polarization signal to
  test whether before this linear to circular polarization conversion
  the linear polarization vectors of the CO maps were aligned with those of the dust.}   
{After the flux correction for the two transitions of the CO spectral
  lines, the new polarization vectors for both CO transitions aligned
  with the dust polarization vectors, establishing that the non-Zeeman
CO circular polarization is due to a linear to circular polarization
conversion. } {}

\maketitle

\section{Introduction}

Magnetic fields have been the subject of many theoretical and
observational studies due to the role they play at different scales in
astrophysics. In molecular clouds and star-forming regions magnetic fields are detected through their effect on the emission of
dust particles and atomic and molecular gas species. For example, in
the submillimeter regime the thermal radiation from non-spherical dust particles becomes linearly
polarized in the presence of a magnetic field.
This emission is polarized perpendicular to the field lines and a map
of such polarization vectors therefore reveals the
plane-of-the-sky orientation of the magnetic field (e.g., \citealt{Hildebrand1999}).  

\subsection{The Goldreich-Kylafis Effect}
Magnetic fields can also cause molecular
spectral lines to be linearly polarized by a few percent
\citep{GK81}. Molecules align with the ambient magnetic field and the presence of a source of
anisotropy in the medium such as velocity gradients parallel or
perpendicular to the magnetic field, or an external
anisotropic radiation causes a population imbalance in the
magnetic sub-levels, $M$. The transitions between these levels
lead to the polarized $\pi$ ($\Delta$$M=0$) and $\sigma$ ($\Delta$$M=\pm1$)
radiation, the former being parallel and the latter perpendicular to
the magnetic field. Depending on which sub-level population dominates,
the detected linear polarization can be in either direction.
This 90$^{\circ}$ ambiguity can in principle be resolved by
obtaining molecular line polarization maps at different transition
frequencies \citep{Cortes05}. 

\subsection{The Zeeman Effect}
Molecular emission can also become circularly polarized through the
Zeeman effect. The local magnetic field removes the energy degeneracy
of the magnetic sub-levels of a molecule, potentially causing a single
spectral line to split into several components. The size of this
splitting is directly proportional to the strength of the magnetic
field. In the case of weak fields, as is the situation in
molecular clouds, a Zeeman line broadening occurs rather than a
splitting, and the line-of-sight component of the field can be
obtained by measuring the net circular polarization of a molecular spectral line. This is so far the only
direct way to measure the strength of interstellar magnetic fields
(e.g., \citealt{Crutcher93, Crutcher99}).

\subsection{This work}

Magnetic field measurements in the submillimeter wavelengths have
mostly been limited to the few aforementioned polarimetry methods, as
well as the ion-neutral line comparison technique of
\citet{Houde2000a} and \citet{li-houde08} 
(see also \citealt{Hezareh10b}), but recent observations have opened new insights into 
physical processes stemming from the interaction between magnetic
fields and interstellar material. \citet{Houde13} reported the
detection of circular polarization in $^{12}$CO ($J=2\rightarrow1$) and
HNCO ($N_{Ka\;Kc}$ = $1_{1,12} \rightarrow 1_{1,11}$) in
Orion KL and proposed a physical model based on anisotropic resonant
scattering in an effort to explain these observations. More precisely, they showed that when the
orientation of the magnetic field changes in the path of propagation
of linearly polarized molecular line radiation, the same species of molecules that emitted this incident
background radiation will resonantly scatter these in the
foreground at their transition frequency. The emerging photons will
acquire a phase shift between their orthogonal scattered components that causes a
transformation of linear into circular polarization. \citet{Houde13}
showed that this can readily account for the levels of circular polarization reported in their observations. 

In this paper, we present the linear polarization maps of $^{12}$CO ($J=1\rightarrow0$)
and ($J=2\rightarrow1$) emission and also of dust emission
obtained near the western molecular edge of the shell of the supernova remnant IC
443. We also show the detection of circular polarization
in both transitions of CO and quantify in our analysis the
amount of conversion of linear to circular polarization and the
consistency of our results with the anisotropic resonant scattering
model. 

We review prior studies performed on IC 443-G in \S 2, present our observations and
data analysis in \S 3, the results in \S 4, a discussion in
\S 5, and end with a short summary in \S 6.

\section{Source Description} \label{sec:sou}
IC 443 is among the most extensively studied
Galactic SNRs, located at a distance of 1.5 kpc
\citep{Fesen84}, in the Gem OB1 association \citep{Heiles84} at
$\alpha\,(J2000)=06^{\mathrm{h}}18^{\mathrm{m}}02\fs7$,
$\delta\,(J2000)=+22^{\circ}39^{'}36^{''}$. IC 443 has an estimated age
of 20,000 yr and reveals a non-circular
morphology with two shells of different radii, namely Shell A at $7$
pc and Shell B at $13$ pc. It is suggested that the
remnant initially evolved in an inhomogeneous medium and that Shell B
may be a part of the remnant blown out into a rarefied
medium \citep{Lee08}. This SNR is
particularly interesting because of the interaction of its so-called Shell A with the
surrounding molecular cloud. The first evidence for this 
interaction was found in shocked OH absorption
lines by \citet{DeNoyer79}. The SNR was mapped in the $J=1\rightarrow0$ transition of CO and
HCO$^+$ emission by \citet{Dickman92}, who identified
eight clumps that were then labeled A to H. The clumps are arranged in a fragmentary,
flattened ring with Clump G (hereafter IC 443-G) being the most
massive of all at approximately $2\times10^3$ M$_{\odot}$ \citep{Xu11}, located in the
westernmost edge of this ring. The correspondence of the HCO$^+$
clumps with the H$_2$ peaks previously identified by \citet{Burton88,
  Burton90}  was an additional sign of interaction of the SNR with its surrounding molecular cloud.

The physics of the shocked gas within Clump G is modeled as a tilted
molecular ring, with the shock transverse to the line of
sight \citep{Dickman92, vanDishoeck93}. However, it has been suggested
that in this clump, a mixture of both J-type and C-type shocks rather
than a single shock model can satisfyingly account for the
observations \citep{Burton90, WS92}. 

Besides the extensive multi-wavelength studies of the SNR structure and 
the physical parameters of the molecular clumps associated with it,
the magnetic field in this source has also been explored to some
degree. Characterizing the direction of the magnetic field
with respect to the structure of the shocks 
is important to constrain multi-dimensional models of such regions, as illustrated in the
shock studies of \citet{Kristensen08} and \citet{Gustafsson10} in the
context of similar molecular shocks associated with star formation
processes. \citet{Wood91} performed polarization observations of
the filamentary structure of the northeast edge of IC 443 at $6$ cm
with the Very Large Array (VLA) and found a correlation between the
strength of the field with regions of relatively high polarized
intensity, but  the direction of the mean field in those regions was
on average oriented neither 
parallel nor perpendicular with respect to the rim.  \citet{Claussen97} performed a high resolution search in the
SNR for OH  maser lines at 1720.53 MHz and found a total of six maser spots, all within
IC 443-G. Masing lines of OH at this frequency typically arise in the
molecular gas behind a slow, non-dissociative
shock \citep{Lockett99}. Although no estimate was obtained for the
line-of-sight component of the magnetic field, the discovery of OH
(1720 MHz) masers associated with the shocked molecular gas supports
the idea that the shock is propagating into a dense magnetized
molecular gas. \citet{Koo10}
measured the Zeeman splitting of the \ion{H}{I} 21 cm
emission line from shocked atomic gas in  
IC 443 and derived an upper limit of 100-150 $\mu$G on the strength of
the line-of-sight field component, which is considerably
smaller than the field strengths expected from a strongly shocked
dense cloud. They concluded that either the
magnetic field within the telescope beam is mostly randomly oriented
or the high-velocity \ion{H}{I} emission is from a shocked interclump medium
of relatively low density. In the most recent radio continuum polarization survey at
6 cm, \citet{Gao11} found the magnetic field around the SNR
dipole-shaped with the configuration being radial around the north
eastern rim. To our knowledge there has been no published higher resolution
studies at shorter wavelengths on mapping the magnetic field in this source and this is what
we accomplish in this work.

\section{Observations and Data Processing}
\subsection{IRAM 30-m Observations}

We obtained on-the-fly maps of CO ($J=1\rightarrow0$) at 115.271 GHz
and CO ($J=2\rightarrow1$) at 230.538 GHz in decent weather conditions ($0.08 <
\tau_{225} < 0.12$) in IC 443-G between 2012 January 6 and 10 with
the IRAM 30-m telescope on Pico Veleta, in the Spanish Sierra
Nevada. We used the EMIR receiver \citep{Carter12}, which is equipped
with four dual-polarization spectral 
bands, each with orthogonal linearly polarized feed horns. Since EMIR allows
for dual band observations with certain band
configurations, we used the E0/E2 combination centered 
at 90 GHz and 230 GHz, respectively, to map the two transitions of CO
simultaneously in the upper sideband of the double-sideband receiver
system. We used the XPOL polarimeter, a special configuration of the
digital correlator backend VESPA, \citep{Thum08} for
the polarization measurements. The Stokes parameters relate to the
time-averaged horizontal ($E_x$) and vertical ($E_y$) electric fields
from the receiver horns and their relative phase difference $\phi$ as follows
\begin{equation}
\begin{array}{ccc}
I &= &\langle\,E^2_x\,\rangle+\langle\,E^2_y\,\rangle\\
\\
Q &=&\langle\,E^2_x\,\rangle-\langle\,E^2_y\,\rangle\\
\\
U &=&2\langle\,E_x\,E_y\,\mathrm{cos}\,\phi\,\rangle\\
\\
V &=&2\langle\,E_x\,E_y\,\mathrm{sin}\,\phi\,\rangle.\\
\end{array}
\label{eq:IQUV}
\end{equation} 

As the above equations imply, the Stokes $I$ and $Q$ are calculated
from total power measurements while Stokes $U$ and
$V$ are obtained through the cross-correlation of the IF signals. All
Stokes parameters are obtained in the Nasmyth reference frame, from
where they are transformed to the equatorial reference frame following
the IAU definition. In this convention, polarization angles increase  
counter-clockwise from sky north,  and the sign of Stokes $V$ is
determined from $V=E^2_{\mathrm{RHC}} - E^2_{\mathrm{LHC}}$, where
RHC (LHC) is the
right (left)-handed circular polarization, with its electric field
vector rotating counter-clockwise (clockwise) as seen by the
observer.

For the two frequency bands E2 and E0, the average line-of-sight system
temperatures were 155 K and 224 K leading to typical antenna
temperature noise of 80 and 100 mK in a resolution bandwidth of 156
KHz, respectively. XPOL was configured with a
channel spacing of 156.25 kHz, covering a bandwidth
of 53 and 105 MHz for the CO ($J=1\rightarrow0$) and ($J=2\rightarrow1$) transitions, respectively. We
obtained 24 on-the-fly maps with a size of $200''\times 140''$ at
reference coordinates $\alpha (J2000)=06^{\mathrm{h}}16^{\mathrm{m}}42\fs21$, $\delta (J2000)=22^{\circ}31' 23\farcs6$ at a
position angle of $17^\circ$ (measured eastward from sky North) in pairs of orthogonal scanning modes. Taking
into account temperature and phase calibrations and off-source reference measurements, the
observing time to complete a pair of maps was about 45 minutes. We
also obtained OTF maps of Mercury on 2012 February 10 to determine the
instrumental polarization mainly caused by the spurious conversion
from Stokes $I$ into the Stokes parameters $Q$, $U$ and $V$, and to
characterize the polarization in the side-lobes of the telescope
beam. Every pair of OTF maps of the planet took about 10
minutes to complete, and the observations were performed with the required continuum
sensitivity as total-power OTF maps, with 280
and 520 MHz bandwidths and spatial resolutions of $21\farcs3$ and
$10\farcs7$ to match the CO ($J=1\rightarrow0$) and
($J=2\rightarrow1$) map resolutions,
respectively. 

\subsection{IRAM 30-m Data Calibration}

A very important step in obtaining the Stokes parameters
  involves the calibration of temperature and phase during the observations. Temperature
  calibration on the signals from the two orthogonal  EMIR receivers are performed using the traditional
hot/cold load technique, i.e., taking subscans on the sky, an
ambient (hot) load, and a cold load. For phase calibration, one has
to correct for the optical and electronic delays, from the
aperture plane of the telescope all the way to the analog-to-digital
converter of the correlator. These delays are different for the
orthogonally linearly polarized components of the incoming radiation
field and therefore modify their intrinsic phase difference,  $\phi$.
This instrumental phase shift occurring between the
temperature calibration unit in the Nasmyth cabin and the correlator
is measured with a wire grid introduced into the beam in front of the
cold load of the calibration unit \citep{Thum08}. The
calibration procedure is performed before starting every OTF map and takes about a minute to
complete. the XPOL calibration software also determines the sign of
the Stokes parameters in agreement with the IAU convention.

\subsection{IRAM 30-m Data processing and correction for instrumental effects} \label{sec:iram_IP}

The data reduction software was developed using the GILDAS\footnote{http://iram.fr/IRAMFR/GILDAS/}
package. 
The CO spectral
lines in both transitions span a wide range of velocities, from about
$-30$ to $20$ km s$^{-1}$ with the broad line wings implying
that the spectra arise from the shocked gas. A self-absorption
feature is seen in the velocity range of about $-7$ to $-2$ km s$^{-1}$ in
the spectra and we therefore created data cubes for the Stokes parameters as well
as the fractional polarization levels,
 by integrating the relevant parameters across the blue- and
red-shifted wings separately to avoid the self-absorption dip and also to
investigate the polarization pattern across the different velocity
ranges.

The next crucial step was to remove the spurious polarization signal from the
Stokes maps of CO. The response of an ideal receiver, i.e., without any
spurious polarization or unwanted mixing of the Stokes parameters, obeys the following set of equations ($*$ denotes the convolution product)
\begin{equation}
S_{\rm i, obs} = B_{\rm i, sky} * P_0,\hspace{20 mm} i=0,1,2,3
\label{eq:noIP}
\end{equation}
where $S_{\rm i, obs}$ is the measured flux density with the indices $i=0$
to 3 refering to the Stokes parameters $I$, $Q$, $U$ and $V$,
respectively, as a function of $B_{\rm i, sky}$  the sky brightness distribution in the
corresponding Stokes parameter and $P_0$ the total power beam. However, the devices in the path
along which the radiation is propagating, i.e., horns, beam splitters, 
the reflector and sub-reflector and other mirrors, can excite electromagnetic modes of
which the net polarization is non-vanishing. The deviation of the
Stokes beams from the total power beam pattern $P_0$ is mainly caused
by the misalignment between the orthogonally polarized horns and also
by the optical elements in the warm and cold optics (e.g.,
elliptical mirrors), and to a lesser degree from non-optimal illumination
of the sub-reflector by the invidual horns. The former leads to Stokes beams that
are invariable with respect to changes in elevation and parallactic angles in the
Nasmyth focus of the telescope, while the latter adds a higher order
elevation dependence. 

As mentioned above, the main source of instrumental polarization in
the observations is the leakage of the signal in the Stokes $I$ into
the other Stokes parameters. A higher order effect is the spurious conversion among the other Stokes
parameters. Since Stokes $Q$ is obtained through auto-correlations
while $U$ and $V$ are calculated through cross-correlations, the exchange of flux between Stokes $Q$ and
the latter two in the Nasmyth cabin is negligible. But after the rotation
of the Stokes $Q$ and $U$ to the sky reference frame, there will be
some mixing between these parameters that may affect the accuracy of
the polarization angles. The latter, however, depends mainly on
the orientation of the calibration grid in the telescope cabin, which has
been measured with a precision of $0.5^{\circ}$ \citep{Thum08}. The polarization angle
calibrations on the Crab nebula \citep{Aumont10} and further tests
using SiO masers \citep{Thum08} show that contributions from receiver
noise and instrumental polarization signals affect the angle calibrations more
strongly than the rotation of the Stokes parameters to the Equatorial frame.

The spurious power exchange between
Stokes $U$ and $V$ happens due to the residual error in the phase
calibration. We can express the observed Stokes $U$ in terms of the 
intrinsic Stokes $U$ and $V$ as 
\begin{equation}
 U_{obs}=U_{int}-V_{int}\,\Delta\phi, 
\label{eq:UvsV}
\end{equation}
where $\Delta\phi$ is the residual phase error (to the first
  order). From earlier polarization measurements at the IRAM 30-m 
on the Crab Nebula, $\Delta\phi$ was measured to be 0.02 radians
across an $11^{\prime\prime}$ beam \citep{Wiesemeyer11}. Therefore, even for comparable
intrinsic Stokes $U$ and $V$ in a measurement, the power leakage from
$U$ to $V$ is about $2\%$ of the intrinsic $V$, which is negligible.

In light of the aforementioned sources of spurious polarization
signals we express  the equations for our Stokes $Q$, $U$ and $V$ measurements with
\begin{equation}
S_{\rm i, obs} = B_{\rm i, sky} * P_0 + B_{\rm 0, sky} * P_{\rm i}
\label{eq:plusIP}
\end{equation}
for $i=1,2,3$, respectively. The second term in the above equation describes
the leakage from Stokes $I$ into the other parameters. The 
instrumental response functions $P_{\rm i}$ can be measured on a
strong and unpolarized source, such as Mercury. 
However, the phase effect of the planet could lead to a
residual linear polarization even in spatially unresolved
observations \citep{Greve09}. We avoided this issue by carrying out the
observations when the planet was almost in its full phase (0.995)\footnote{The physical ephemeris data are taken from the {\it Astronomical Almanac 2012},
table E56.} and as compact as $\Theta_{\mathrm{M}} = 4\farcs9$, i.e., spatially
unresolved even in the $10\farcs7$ beam of the CO ($J=2\rightarrow1$)
data. The Stokes maps of Mercury  may be represented as 
\begin{equation}
S_{\rm i,\mathrm{M}}=B_{\rm 0, \mathrm{M}} * P_{\rm i}, 
\label{eq:mercury}
\end{equation}
where $B_{\rm 0, \mathrm{M}}$ is Mercury's intrinsic brightness
distribution approximated by a disk whose constant surface intensity is
normalized to unity. We define the function $G$, such that $G=B_{0, \mathrm{M}}
* P_0$ and approximate it as a Gaussian with a full-width-half-maximum (FWHM) of 
\begin{equation}
\Theta = \sqrt{\Theta_{\mbox{\sc fwhm}}^2+\frac{\ln{2}}{2}\Theta_{\mathrm{M}}^2},
\label{eq:Gwidth}
\end{equation}
where $\Theta_{\mbox{\sc fwhm}}^2$ is the width of the telescope beam
 $P_0$. The FWHM of this Gaussian function is $21\farcs 5$ for 
the CO ($J=1\rightarrow0$) maps and $11\farcs 1$ for the
CO ($J=2\rightarrow1$) maps. We convolve this Gaussian function with the images from
 Equation (\ref{eq:plusIP}) to obtain
\begin{equation}
S_{\rm i, obs} * G = B_{\rm i, sky} * P_0 *G + S_{\rm 0, obs} * S_{\rm
  i, \mathrm{M}},  
\label{eq:IPcorr}
\end{equation}
where the second term from the right, $S_{\rm 0, obs} * S_{\rm
  i, \mathrm{M}}$, is the convolution of Stokes $I$ map of IC 443-G with
each of the Stokes maps of Mercury and is our model for the instrumental
polarization. We subtract these convolution products from the corresponding Stokes maps of our
target (the left hand side of Equation (\ref{eq:IPcorr})) to obtain images of the instrinsic sky brightness distribution of the
Stokes parameters $Q$, $U$ and $V$ (Equation (\ref{eq:noIP})) but with degraded spatial
resolutions of $30\farcs 3$ and $15\farcs 4$ for the
CO ($J=1\rightarrow0$) and ($J=2\rightarrow1$) transitions,
respectively. Finally, the reduced maps were deconvolved to the original
resolutions of the respective CO beams. 

\begin{figure*}
\begin{center}  
\begin{tabular}{cc}  
\includegraphics[angle=-90,scale=0.4]{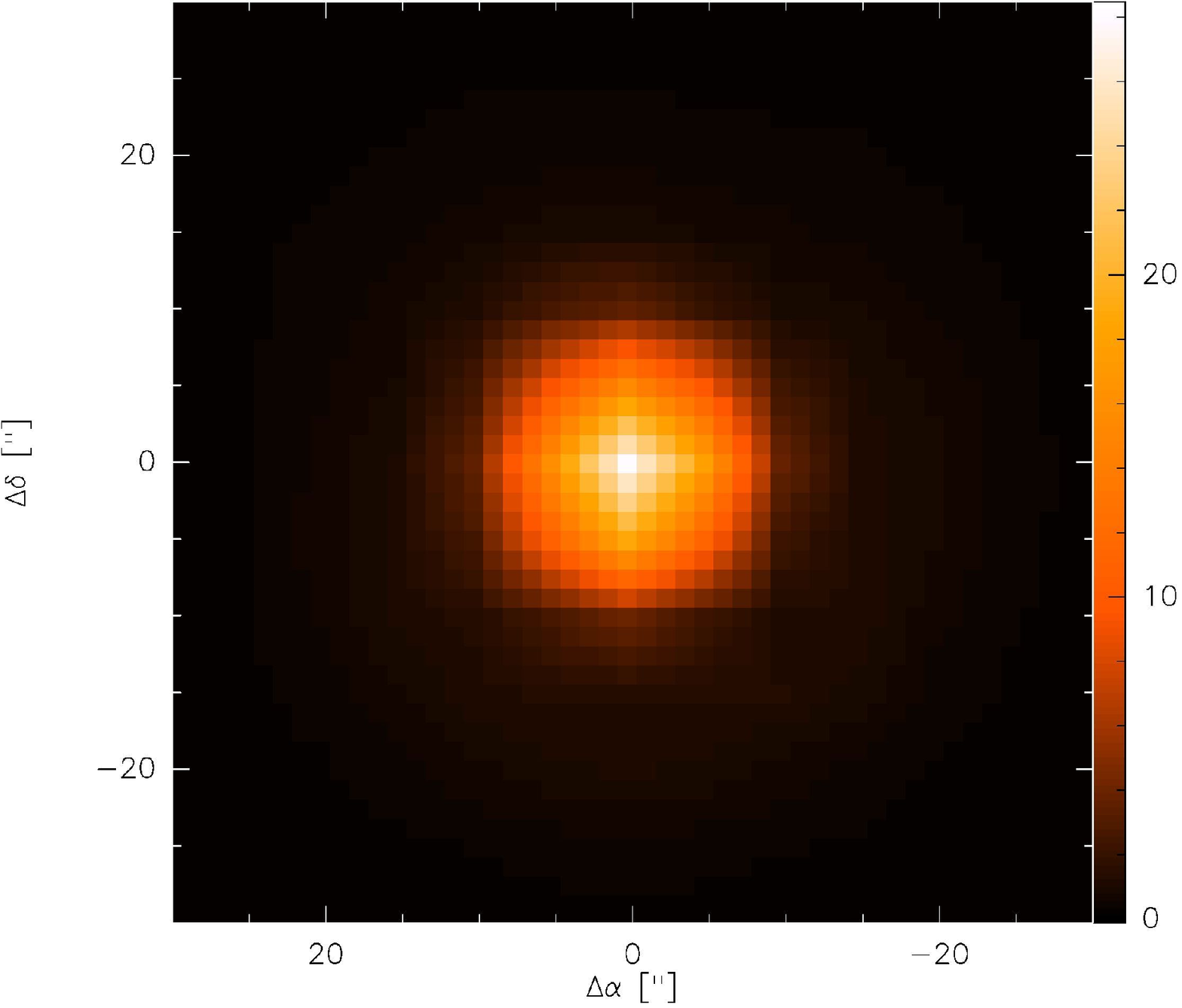}
& \includegraphics[angle=-90,scale=0.4]{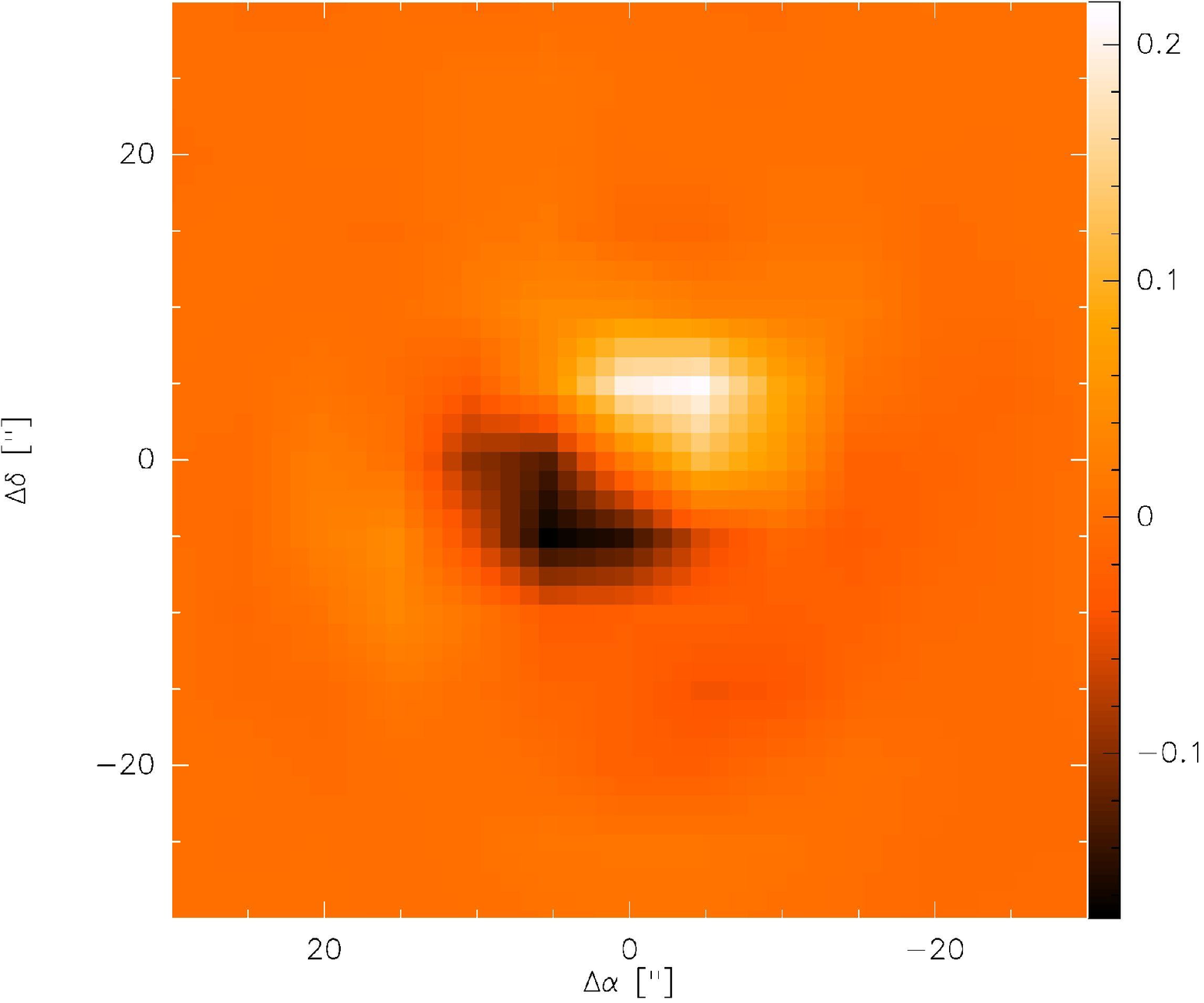} \\ 
\includegraphics[angle=-90,scale=0.4]{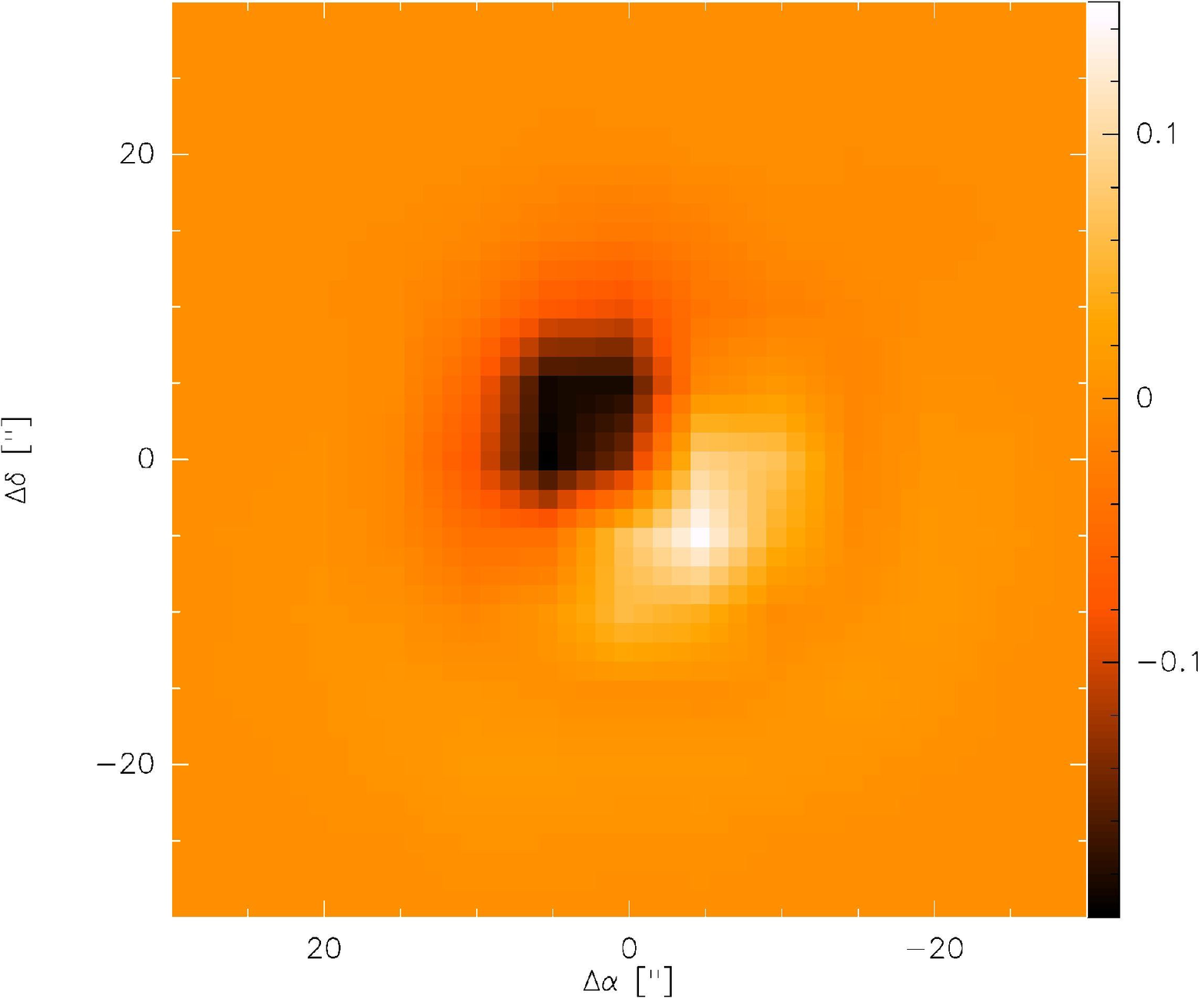}
& \includegraphics[angle=-90,scale=0.4]{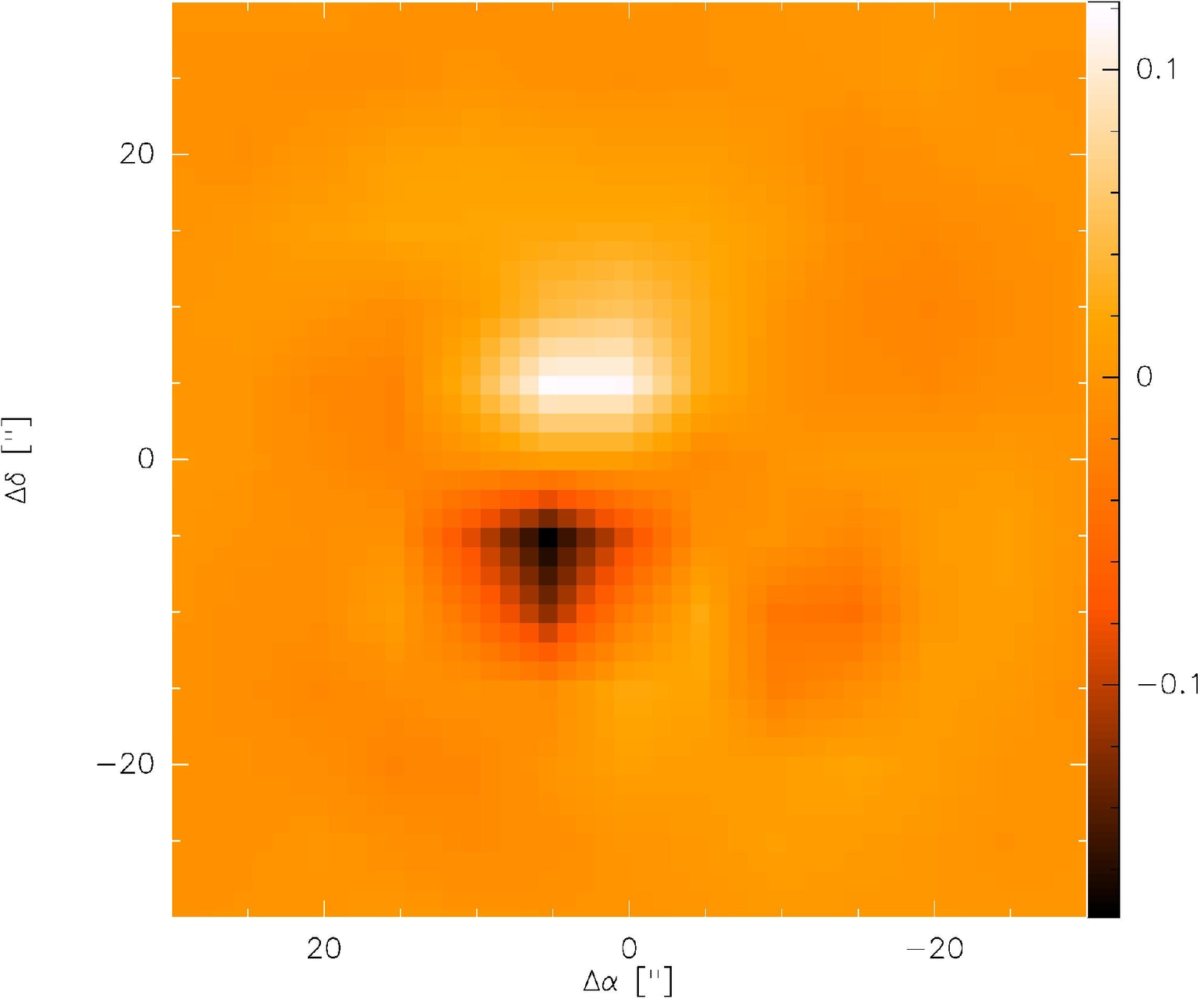} \\       
\end{tabular}   
\end{center}  
\caption{Starting from top left, going clockwise are the intensity maps of Mercury in Stokes $I$, $Q$, $U$, and
  $V$ at $1.3$ mm. The color wedge on the right vertical scale of the
  figures has units of Jy/beam and the maps are over-sampled by a
  factor of 4 for better display. } 
\label{fig:mercury1mm} 
\end{figure*}
\begin{figure*}
\begin{center}  
\begin{tabular}{cc}  
\includegraphics[angle=-90,scale=0.4]{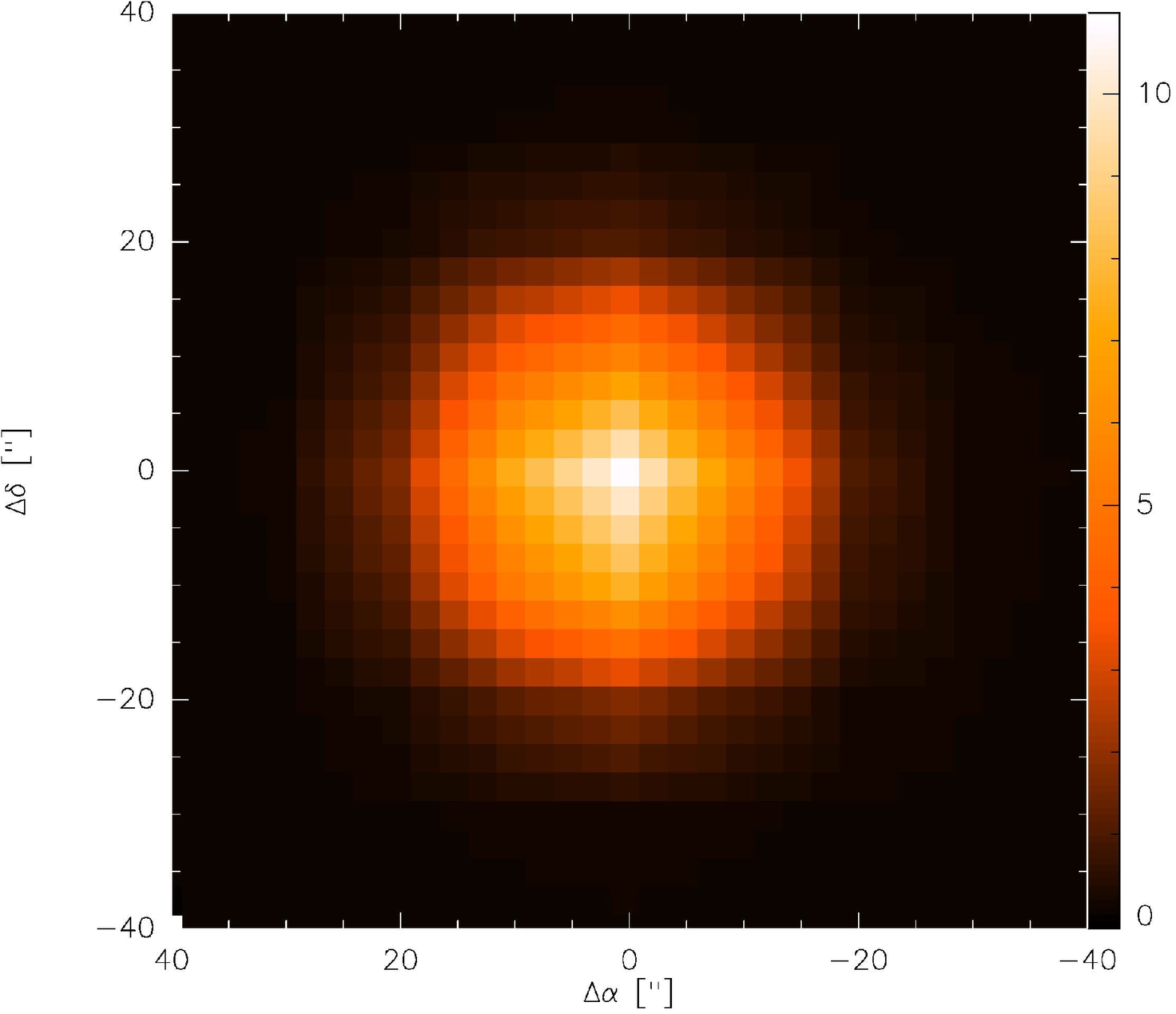}
& \includegraphics[angle=-90,scale=0.4]{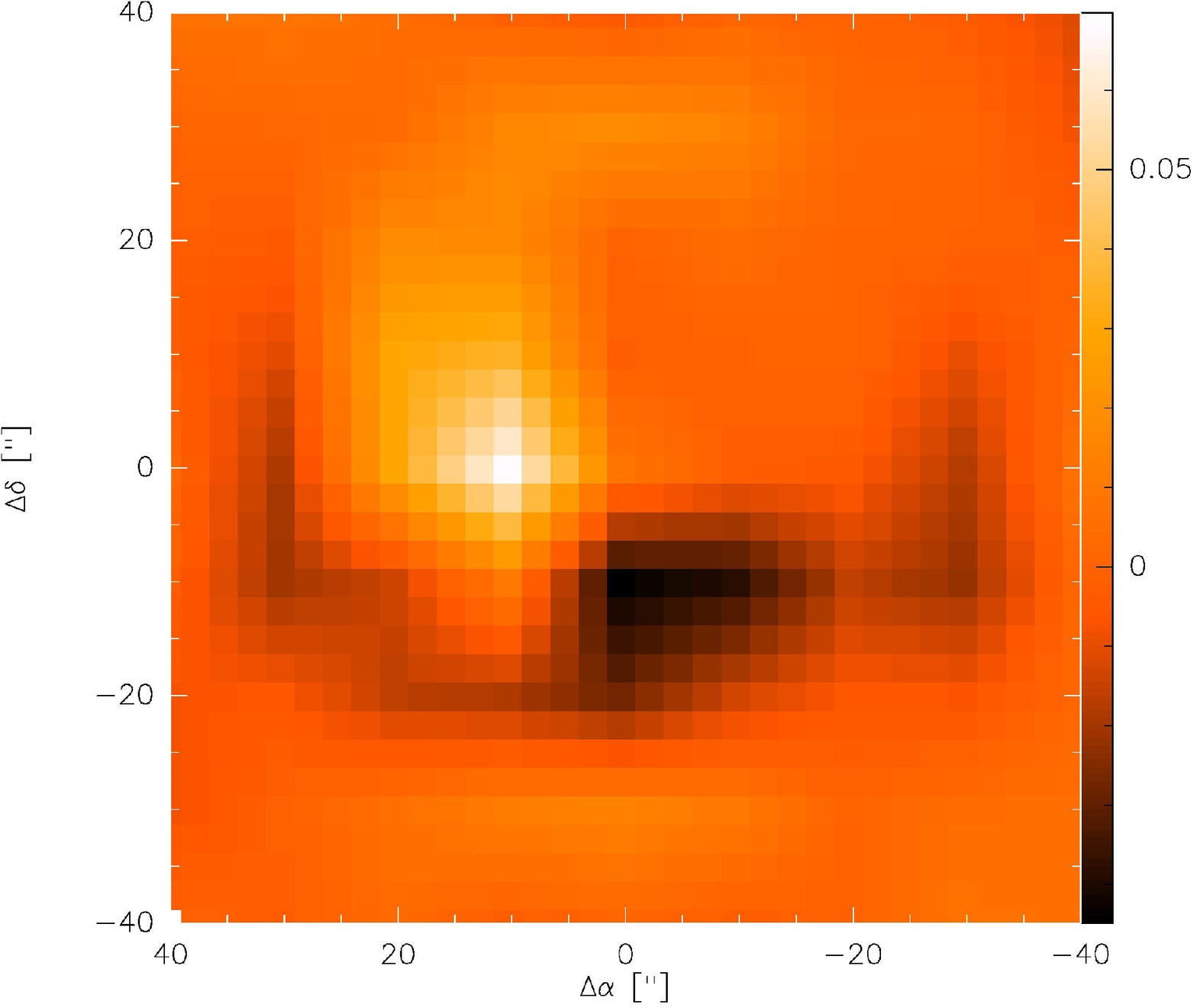} \\ 
\includegraphics[angle=-90,scale=0.4]{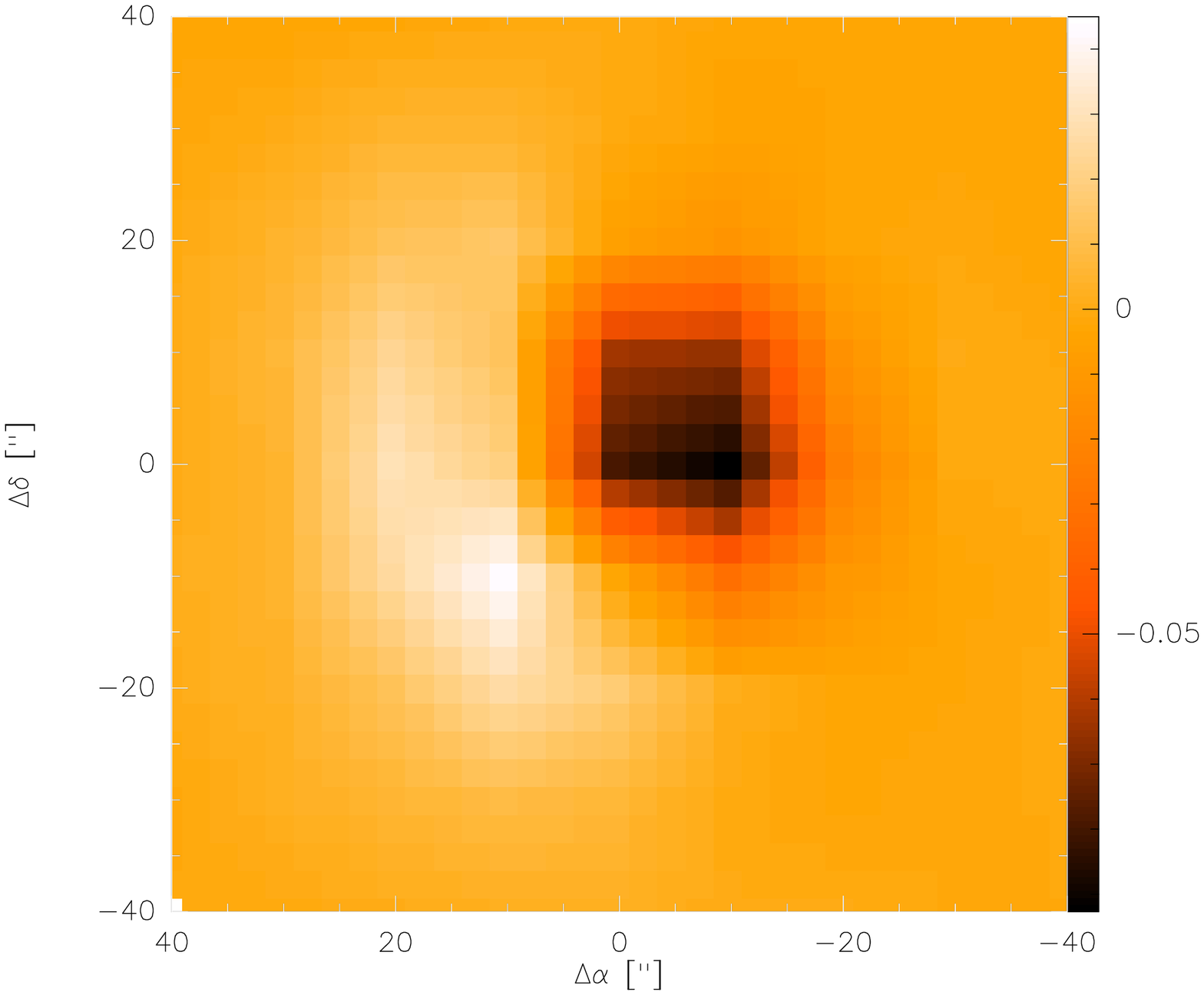}
& \includegraphics[angle=-90,scale=0.4]{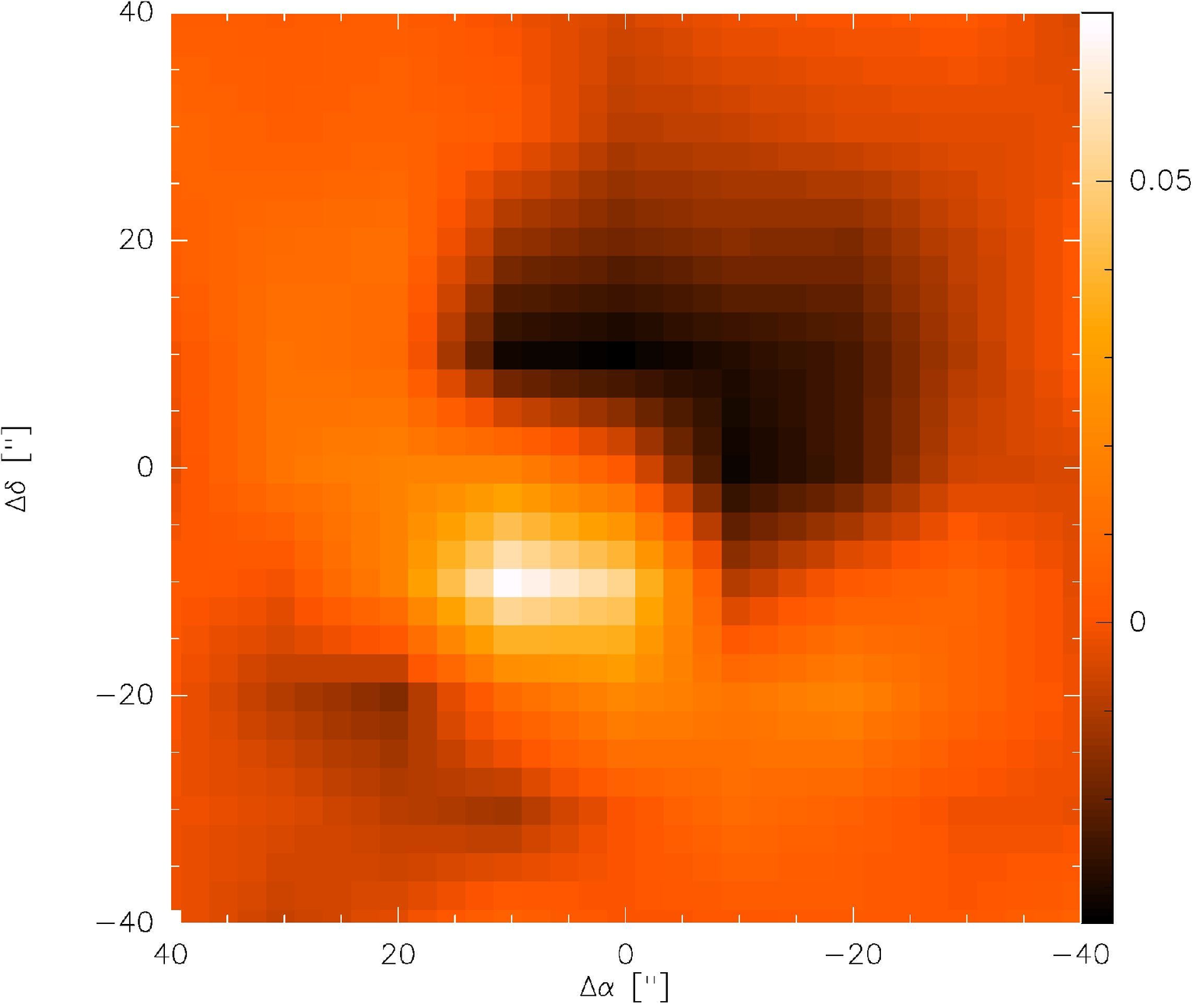} \\       
\end{tabular}   
\end{center}  
\caption{Same order of maps as in Figure \ref{fig:mercury1mm} are the intensity
  maps of Mercury in Stokes $I$, $Q$, $U$, and $V$ at $3$ mm. These
  maps are also over-sampled by a factor of 4 for better display.} 
\label{fig:mercury3mm} 
\end{figure*}

Figure \ref{fig:mercury1mm} shows
the Stokes maps for Mercury at 1 mm while Figure \ref{fig:mercury3mm}
shows the Stokes maps at 3 mm. The Stokes maps of Mercury 
are over-sampled by a factor of four for better display purposes. The amount of instrumental polarization that we
measured towards Mercury on the optical axis were $Q/I=0.2$~\%,
$U/I=0.09$~\% and $V/I=0.5$~\% at 1 mm, with the side-lobes reaching $1.2\%$ in the reference frame of the Nasmyth
focus. The corresponding polarization levels at 3 mm were $Q/I=0.1$~\%,
$U/I=0.02$~\% and $V/I=0.6$~\% on the optical axis and $0.8\%$ in the side-lobes.

In order to subtract the instrumental polarization from the data
correctly, we had to rotate the Stokes maps of Mercury, which
  are invariable in the Nasmyth reference frame, to the same elevation
  and parallactic angles of the IC 443-G maps in the plane of the sky:

\begin{equation}
 \begin{array}{ccc}
\Delta\alpha_{\mathrm{IC\,443-G}}& = &
\Delta\alpha_{\mathrm{M}}\,\mathrm{cos}(\Delta\gamma)+\Delta\delta_{\mathrm{M}}\,\mathrm{sin}(\Delta\gamma)\\ 
\\
\Delta\delta_{\mathrm{IC\,443-G}}& = &
-\Delta\alpha_{\mathrm{M}}\,\mathrm{sin}(\Delta\gamma)+\Delta\delta_{\mathrm{M}}\,\mathrm{cos}(\Delta\gamma).
\end{array}
\label{eq:delta_ad}
\end{equation}

Here $\Delta\alpha$ and $\Delta\delta$ are the coordinate
  offsets from the map center, $\Delta\gamma
  =\gamma_{\mathrm{M}}-\gamma_{\mathrm{IC\,443-G}}$, where
  $\gamma=\epsilon-\eta$, with $\epsilon$ the elevation and $\eta$ the
  parallactic angle of the sources. The Stokes $Q$ and $U$ of each
  source was transformed from the Nasmyth reference frame 
to the equatorial reference frame with the aforementioned IAU definition of $Q$ and $U$ as

\begin{equation}
\begin{array}{ccc}
Q_{\mathrm{eq}} &=& -Q_\mathrm{N}
\mathrm{cos}(2\gamma)-U_\mathrm{N}
\mathrm{sin}(2\gamma)\\
\\
U_{\mathrm{eq}} &=& Q_\mathrm{N} \mathrm{sin}(2\gamma)-U_\mathrm{N} \mathrm{cos}(2\gamma)
\end{array}
\label{eq:QU_eq}
\end{equation}

In this correction algorithm, however, we assume that the
instrumental Stokes $Q$ and $U$ are stationary during the OTF mapping
of IC 443-G. Indeed, we examined the impact of this assumption by
comparing the values for $\gamma$ recorded in every pair of OTF maps, between the beginning of the
first and the end of the second map.
The difference between the two $\gamma$ angles range from $3^{\circ}$
to $34^{\circ}$, or $16^{\circ}$ on average (the maximum is only reached
once per day, 
namely, during transit). The removal of the instrumental polarization
as described above yielded results that are indistinguishable within
$1\sigma$ limits of the instrumental polarization levels. 

\subsection{APEX Observations}

Besides the CO polarization observations at the IRAM 30-m telescope,
we observed IC 443-G with the continuum polarimeter at the APEX
telescope \citep{Guesten06}, named 
PolKa after the German POLarimeter f\"{u}r bolometer KAmeras
\citep{Siringo04}. PolKa works in combination
with LABOCA, the Large APEX Bolometer Camera \citep{Siringo09} and
uses a reflection-type rotating half-wave plate (RHWP) that functions as the
polarization modulator. The RHWP modulates the polarized component of
the signal at $4$ times its angular speed. For a complete description
of the design and operation of PolKa, see \citet{Siringo04, Siringo12}.

Continuum observations of IC 443-G were performed at 345 GHz within a
bandwidth of 60 GHz in December 2011 and November
2012, for a total of 4.2 hours on-source integration time with average PWV values of $0.8$ mm, or
alternatively $\tau_{225}\simeq0.035$. We used OTF mapping in compact raster-spiral mode that produces
uniform coverage maps over the size of the field of view of LABOCA
($11'\times11'$, \citealt{Siringo09}) and set the modulation
frequency to $6.24$ Hz, safely above the typical frequencies of
atmospheric fluctuations. Polarization observations at APEX are
  performed at the Cassegrain focus and the parallactic angle 
is corrected with negligible errors, via conversion from
the horizontal to the equatorial frame, in the same manner as
for regular LABOCA observations.

\subsection{APEX Data Analysis}

\subsubsection{Demodulation of the Polarized Signal} \label{sec:demod}

The bolometer detectors at APEX are not 
polarization sensitive, therefore for polarization observations a
linear polarizer was required in the
beam path  to produce an intensity modulation that would be
recordable by the detectors. The modulated bolometer signal, in the two cases of horizontal (H) and
vertical (V) positions of the analyzer, may be expressed as \citep{Siringo04} 

\begin{equation}
 \begin{array}{ccc}
S_{\mathrm{H}} & = & \frac{1}{2}[I+Q\, \mathrm{cos}\,(4\omega t)+U\, \mathrm{sin}\,(4\omega t)]\\
\\
S_{\mathrm{V}} & = & \frac{1}{2}[I-Q\, \mathrm{cos}\,(4\omega t)-U\, \mathrm{sin}\,(4\omega t)]
\end{array}
\label{eq:polkaS}
\end{equation}
where $\omega$ is the angular speed of the wave-plate. After
absolute calibration of the position angles of the RHWP the Stokes parameters
are extracted using a generalized synchronous demodulation method, explained in
more detail by \citet{Wiesemeyer13}. 

While the RHWP modulates the Stokes $Q$ and $U$
parameters at four times the angular speed, the Stokes $I$ is ideally not expected to
be modulated (Equation (\ref{eq:polkaS})). It was noticed, however, that the
bolometer signals also carry a spurious 
modulation proportional to the total intensity, which
could be a result of standing waves in the system. 
This signal was harmonic and could be removed from
the data by applying a Fourier filter to the data in every scan, while
preserving the modulated polarization signal.  
The modulated signal was further reduced using a
synchronous demodulation method (\citealt{Siringo04,Wiesemeyer13}
).

\subsubsection{Calibration of the Total Power and Linear Polarization}
The calibration of Stokes $I$ was performed by observing primary and secondary
flux calibrators as used for regular bolometer observations. The calibration factor, however, differs
by about a factor of $2$ from the usual LABOCA calibration factor
because of the $50\%$ flux loss due to the insertion of the analyzing
polarizers. We estimate that the calibration is the
same for the two analyzers (vertical and horizontal) within measurement errors. 
The polarized flux and fractional polarization levels were calibrated
by observing the Crab nebula, a well established calibration source
\citep{Aumont10} and further consistency checks where done towards OMC-1.
We note that once the total power and the polarization level of the incoming signal are calibrated
and the instrumental polarization characterized and subtracted from
the data, there is no need to further calibrate the polarization
angles, since the angles are directly calculated from the reduced values
of Stokes $Q$ and $U$ 
\begin{equation}
  \chi=\frac{1}{2}tan^{-1}(\frac{U}{Q}).
\label{eq:angle}
\end{equation}
 
\subsubsection{Instrumental Polarization}

In order to evaluate the level of spurious
polarization in our data we observed the planet Uranus, which can be
considered as an unpolarized point source. We observed Uranus in a linear on-the-fly
mode and measured an upper limit of $0.1\%$ linear polarization towards the center of the
planet's disk. However, the polarization level around the edges is
as high as $3\%$ and is a result of weak Stokes $I$ flux and contribution to
$Q$ and $U$ from the side-lobes. The Stokes maps of Uranus were
convolved with the maps of IC 443-G via a similar method as explained
for the IRAM 30-m data (\S \ref{sec:iram_IP}) to account for the effect of the polarized
side-lobes on the polarization data.

\section{Results}

\subsection{Linear polarization maps of CO}

\begin{figure*}
\begin{center}  
\begin{tabular}{cc}  
\includegraphics[trim = 0mm 0mm 6.5mm 7.5mm, clip, angle=0,scale=0.4]{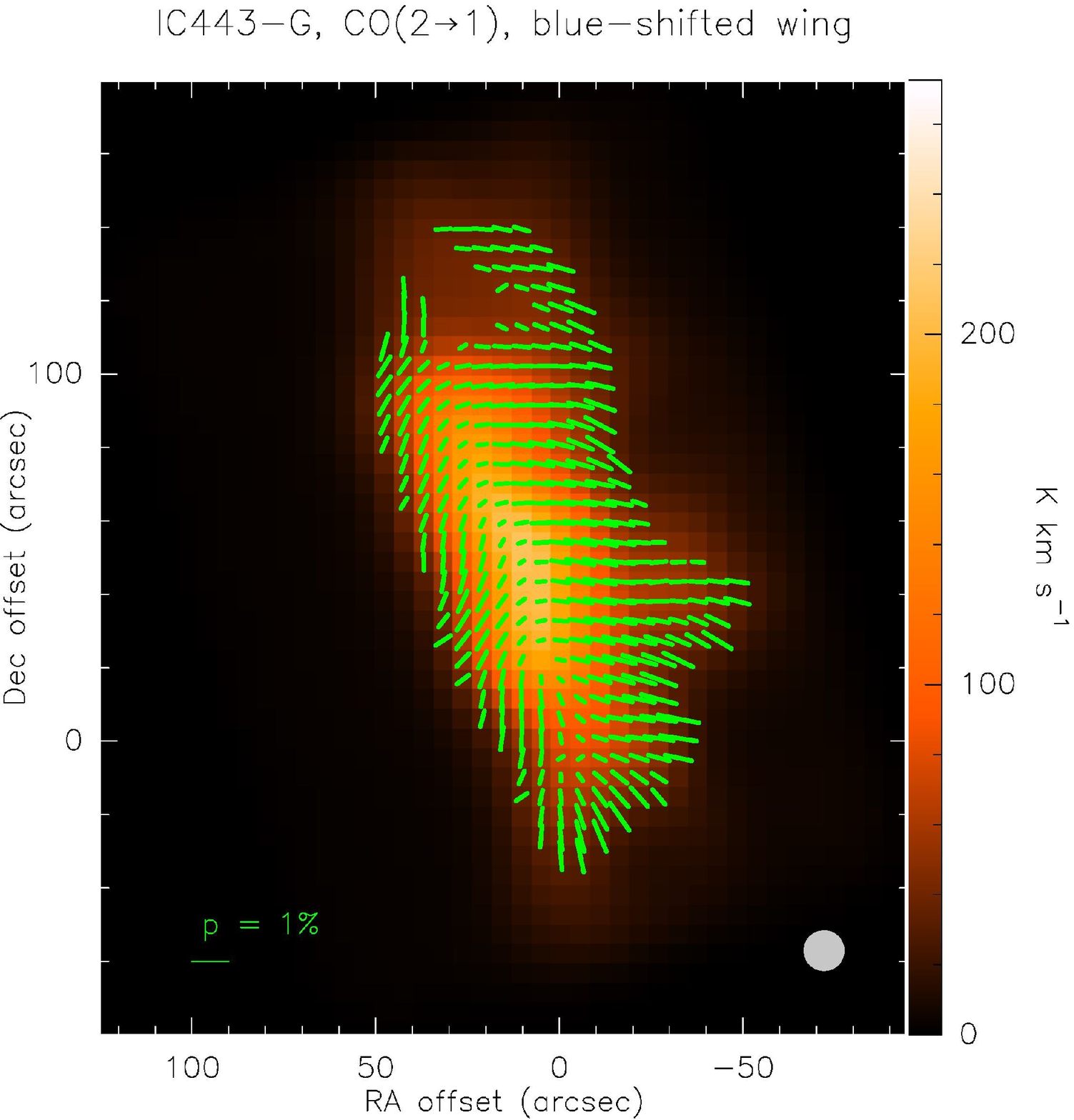}
& \includegraphics[trim = 6.5mm 0mm 0mm 7.5mm, clip, angle=0,scale=0.4]{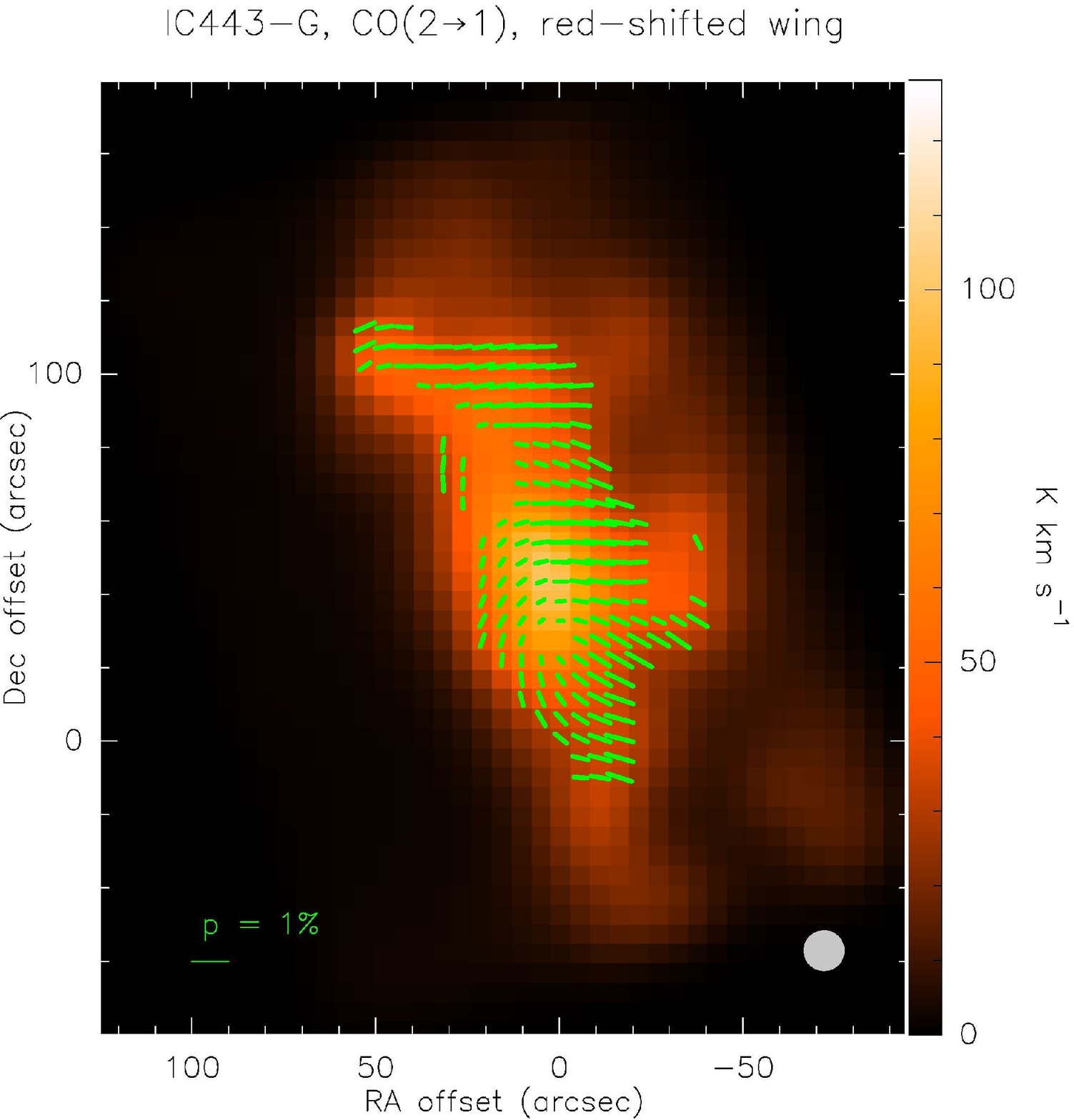} \\ 
\end{tabular}   
\end{center}  
\caption{The polarization maps of CO ($J=2\rightarrow1$) spectral lines
in the blue-shifted ($-30$ to $-7$ km s$^{-1}$, left) and red-shifted
($-2$ to $20$ km s$^{-1}$, right) wings, with quarter-beam sampling. For each
spectral line, the linear polarization levels and angles are calculated across velocity ranges
spanning each line wing separately and only polarization vectors with $p\geq3\sigma_p$
are plotted. These maps are corrected for instrumental polarization and the intrinsic linear polarization levels vary
between $0.5-0.8\%$ across the maps. The beam size is shown in the
lower right corners and the offsets are with respect to the reference coordinates $\alpha\,(J2000)=06^{\mathrm{h}}18^{\mathrm{m}}02\fs7$,
$\delta\,(J2000)=+22^{\circ}39^{'}36^{''}$.} 
\label{fig:CO21} 
\end{figure*}

\begin{figure*}
\begin{center}  
\begin{tabular}{cc}  
\includegraphics[trim = 0mm 0mm 6.5mm 7.5mm, clip, angle=0,scale=0.4]{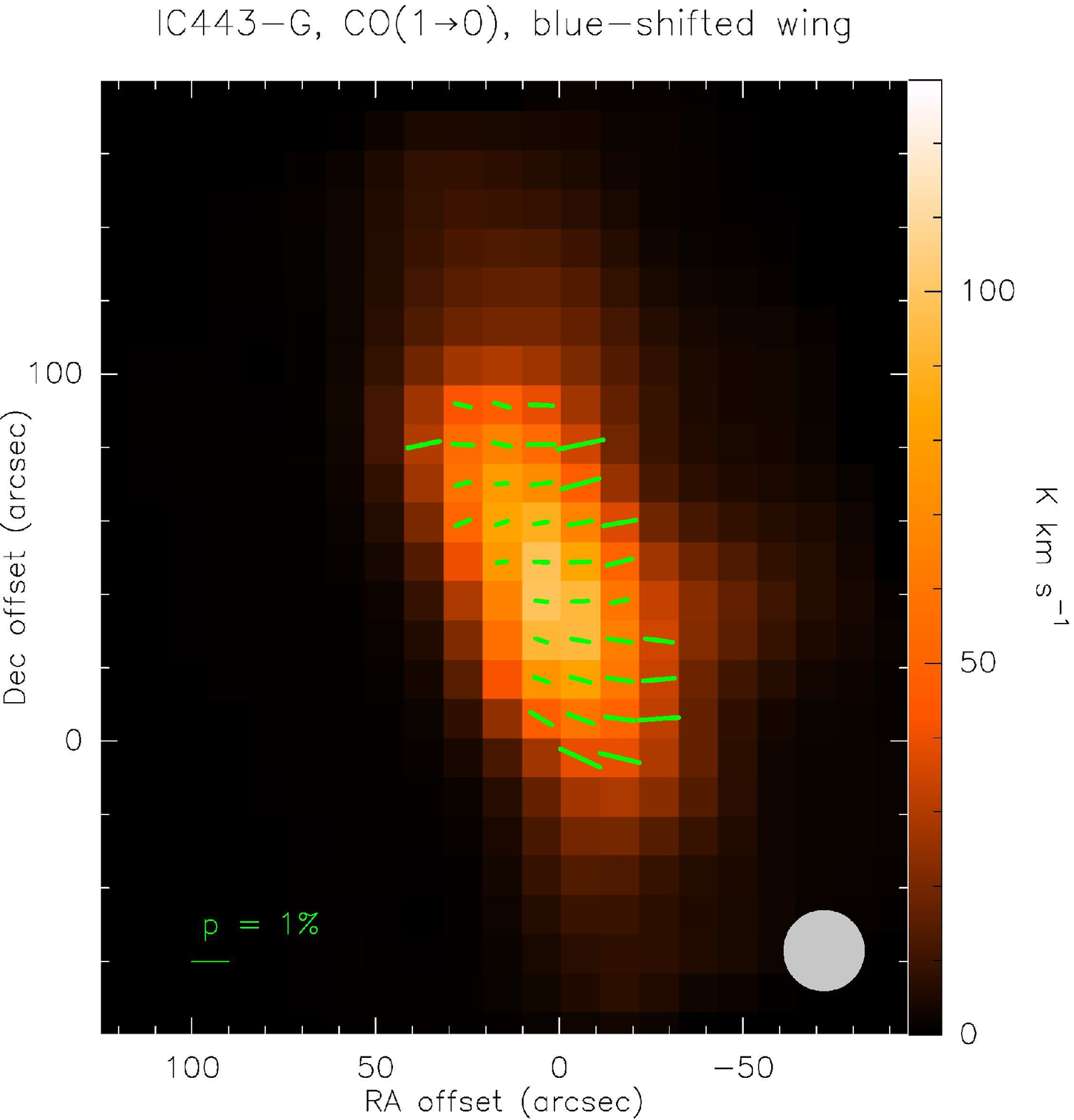}
& \includegraphics[trim = 6.5mm 0mm 0mm 7.5mm, clip, angle=0,scale=0.4]{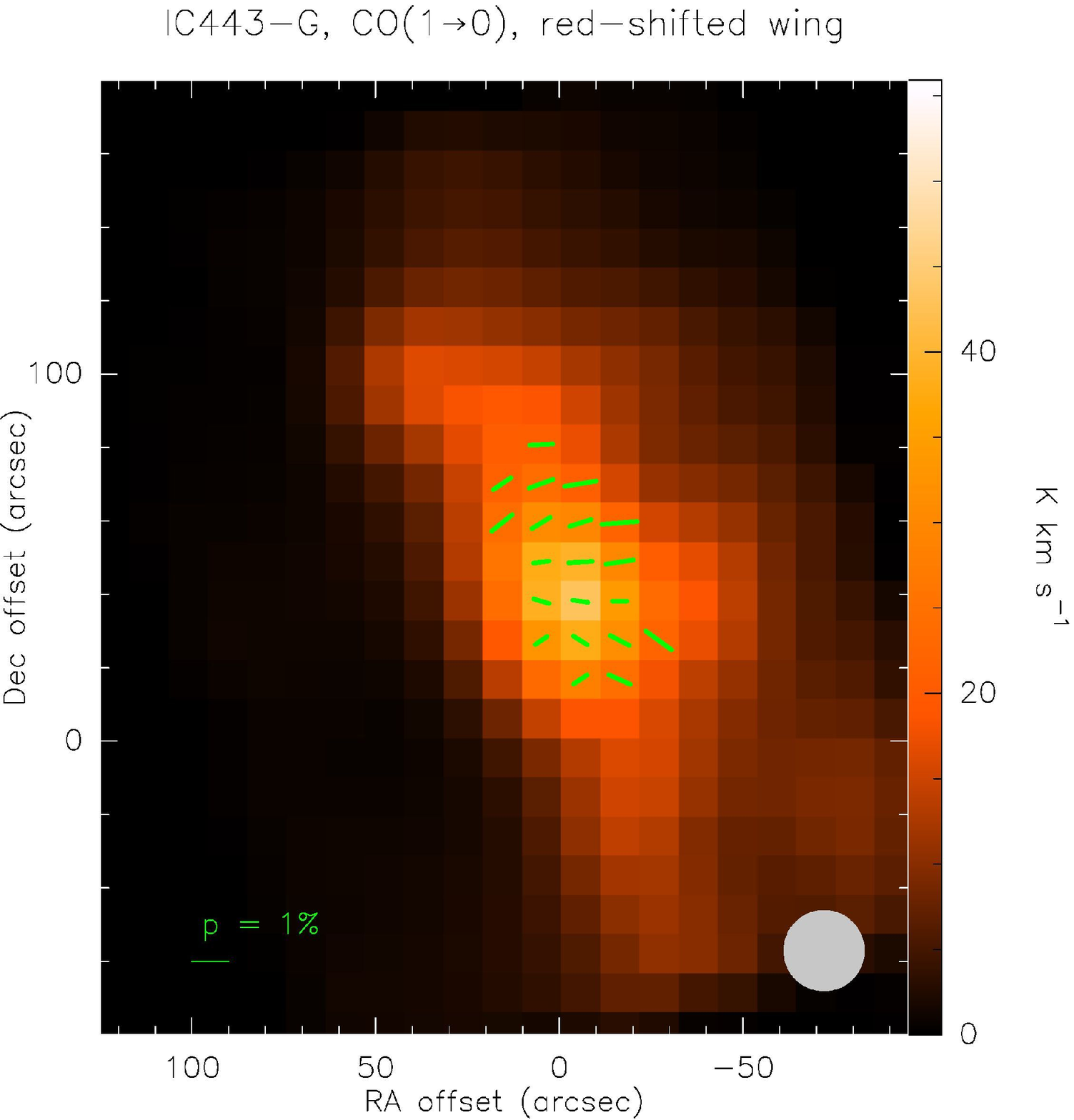} \\ 
\end{tabular}   
\end{center}  
\caption{Same as Figure \ref{fig:CO21} but for the CO ($J=1\rightarrow0$) transition.} 
\label{fig:CO10} 
\end{figure*}

The linear polarization map corresponding to the blue-shifted line wings of CO
 ($J=2\rightarrow1$)  in IC 443-G is displayed in the left
panel of Figure \ref{fig:CO21} and that of the
red-shifted wings is plotted in the right panel. The polarization
level and angle for every given pixel are averaged over a velocity
range of $-30$ to $-7$  km s$^{-1}$ for the blue-shifted wing and $-2$
to $20$ km s$^{-1}$ for the red shifted wing. This way the
self-absorption dip spanning the velocity range of $-7$ to $-2$ km
s$^{-1}$ is avoided in the polarization calculations. The instrumental
polarization is removed from both maps and the polarization vectors
are displayed for pixels with $p\geq3\sigma_{p}$,
where $p$ and $\sigma_{p}$ are the linear
polarization level and its uncertainty, respectively. The Stokes $Q$ and $U$ spectra are calculated in the equatorial frame
and  have  similar r.m.s. noise levels of $82$ mK, corresponding to an
uncertainty of polarization angles $\sigma_{\chi}<9\fdg5$. With
similar polarization uncertainty values, the polarization maps for the CO
 ($J=1\rightarrow0$) emission are presented in Figure \ref{fig:CO10}. 

For a more detailed view of the linear polarization behaviour across a spectral line,
Figure \ref{fig:CO_pL} presents  sample linear polarization
profiles of the CO ($J=2\rightarrow1$) and ($J=1\rightarrow0$) spectral
lines at the peak position of CO emission at offsets ($20^{\prime\prime}$,
$60^{\prime\prime}$) from the reference coordinates.  The Stokes $I$
spectra are plotted on the top panels with
the fractional linear polarization levels, with CO ($J=2\rightarrow1$) smoothed to a resolution of $1.2$
km s$^{-1}$ and CO ($J=1\rightarrow0$) to $1.4$ km s$^{-1}$. The dashed profiles are the Stokes $I$ spectra with original
resolution of $0.2$ km s$^{-1}$ for CO ($J=2\rightarrow1$) and $0.4$ km
s$^{-1}$ for CO ($J=1\rightarrow0$). The interaction of the SNR shock with the
cloud is evident from the broadened wings of the spectral
lines in both transitions. The filled circles show fractional linear polarization levels with $p\geq3\sigma_{p}$
while the empty circles represent polarization levels with
$P+2\sigma_{p}\leq1\%$. The polarization angles $\chi$ corresponding
to the aforementioned fractional polarization levels are shown in
the middle panels. Although these spectra show a relatively small
number of polarization measurements with $p \geq 3 \sigma_{p}$ at the
corresponding velocity resolutions, this criterion is easily met when
the signals are integrated over the larger velocity ranges used for
the maps in Figures \ref{fig:CO21} and
\ref{fig:CO10}. Finally, the linear polarization profiles,
$p_L=\sqrt{Q^2+U^2}$, are plotted in the bottom panels. No positive noise
bias correction is applied to these profiles and the data are plotted
without correction for telescope efficiency. The linear polarization
signals measured for the two CO transitions are consistent with what
is expected from the Goldreich-Kylafis effect \citep{GK81}.


\subsection{Circular Polarization in CO Spectral lines}

\begin{figure*}
\begin{center}  
\begin{tabular}{cc}  
\includegraphics[angle=0,scale=0.4]{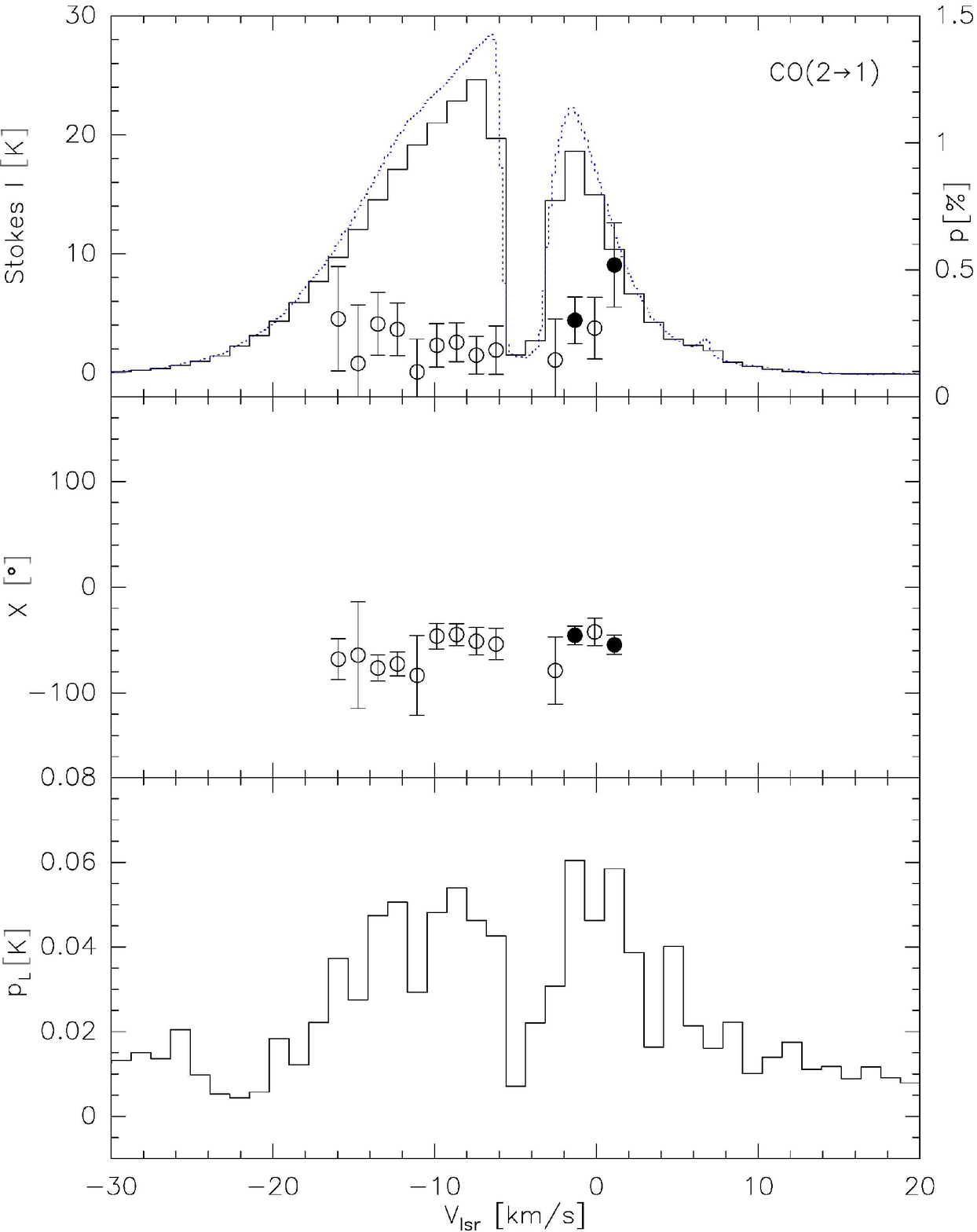}
& \includegraphics[angle=0,scale=0.4]{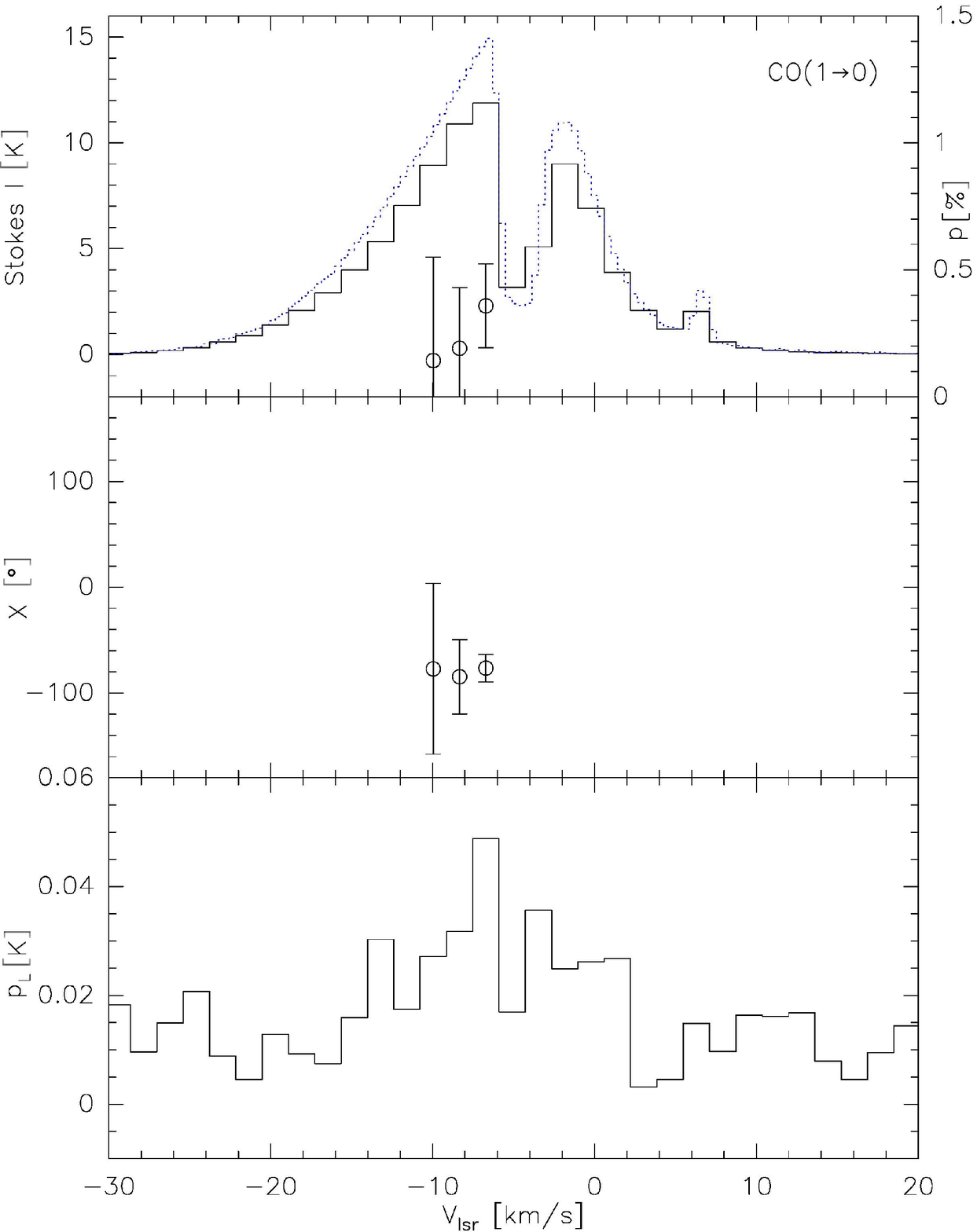}\\ 
\end{tabular}   
\end{center}  
\caption{The linear polarization profiles for
  $^{12}$CO ($J=2\rightarrow1$) (left) and ($J=1\rightarrow0$)
  (right). The top panels show Stokes $I$ spectra at the peak position of  $^{12}$CO emission in
IC 443-G, i.e., at offsets ($20^{\prime\prime}$, $60^{\prime\prime}$) from the map
origin ($\alpha\,(J2000)=06^{\mathrm{h}}18^{\mathrm{m}}02\fs7$,
$\delta\,(J2000)=+22^{\circ}39^{'}36^{''}$), overlaid with fractional linear polarization levels $p$
smoothed to 1.2 km s$^{-1}$ and 1.4 km s$^{-1}$, respectively. The dashed profiles are Stokes $I$ at the original
resolution of 0.2 km s$^{-1}$ and 0.4 km s$^{-1}$ for the higher and
lower transitions, respectively. The
corresponding polarization angles  ($\chi$) are plotted in the middle
panel and the linear polarization ($p_\mathrm{L}=\sqrt{Q^2+U^2}$)
intensity profiles across the spectral line in the bottom. These spectra are
corrected for instrumental polarization but not for positive bias due
to noise.}
\label{fig:CO_pL} 
\end{figure*}

\begin{figure*}
\begin{center}  
\begin{tabular}{cc}  
\includegraphics[angle=0,scale=0.4]{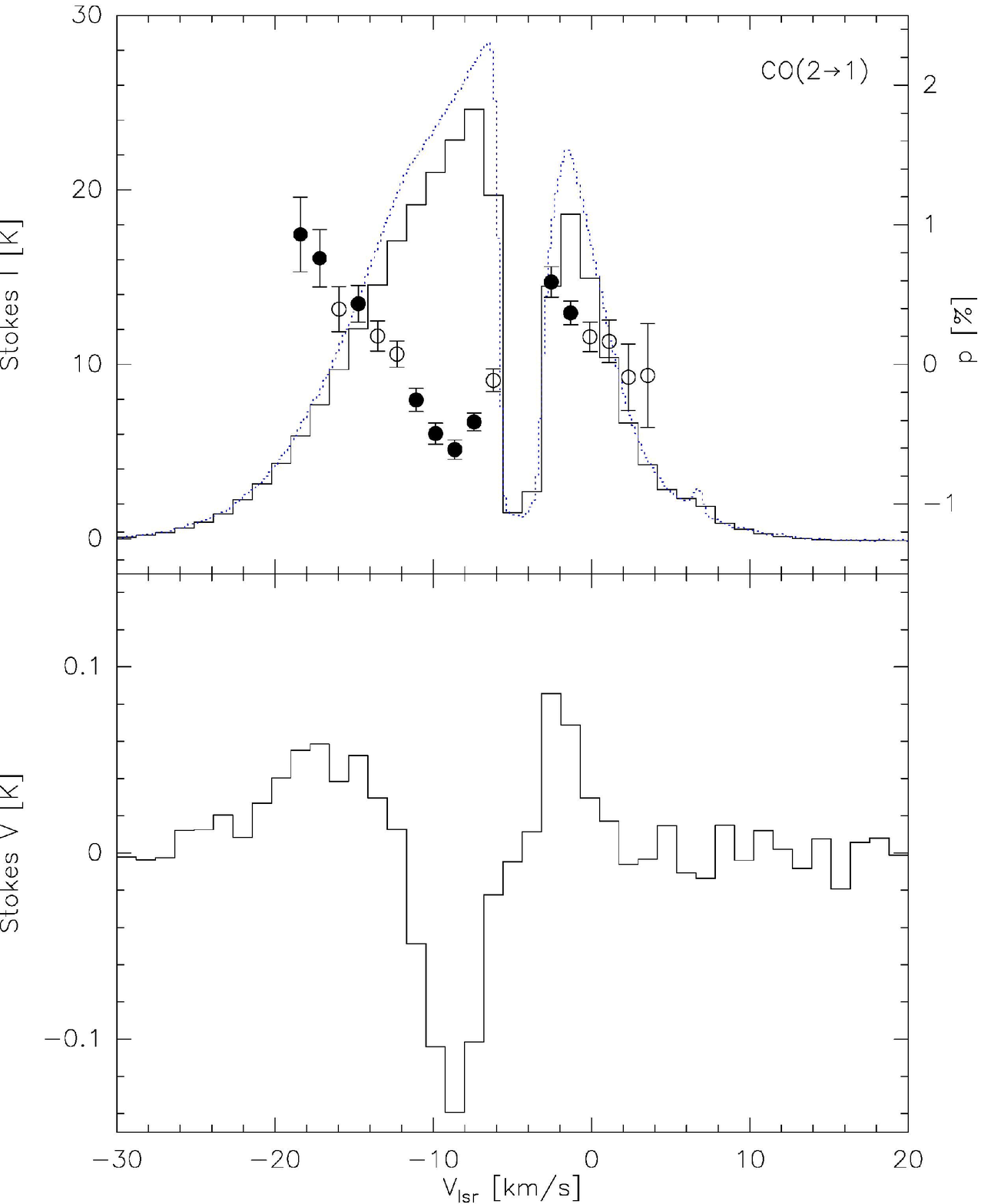}
& \includegraphics[angle=0,scale=0.4]{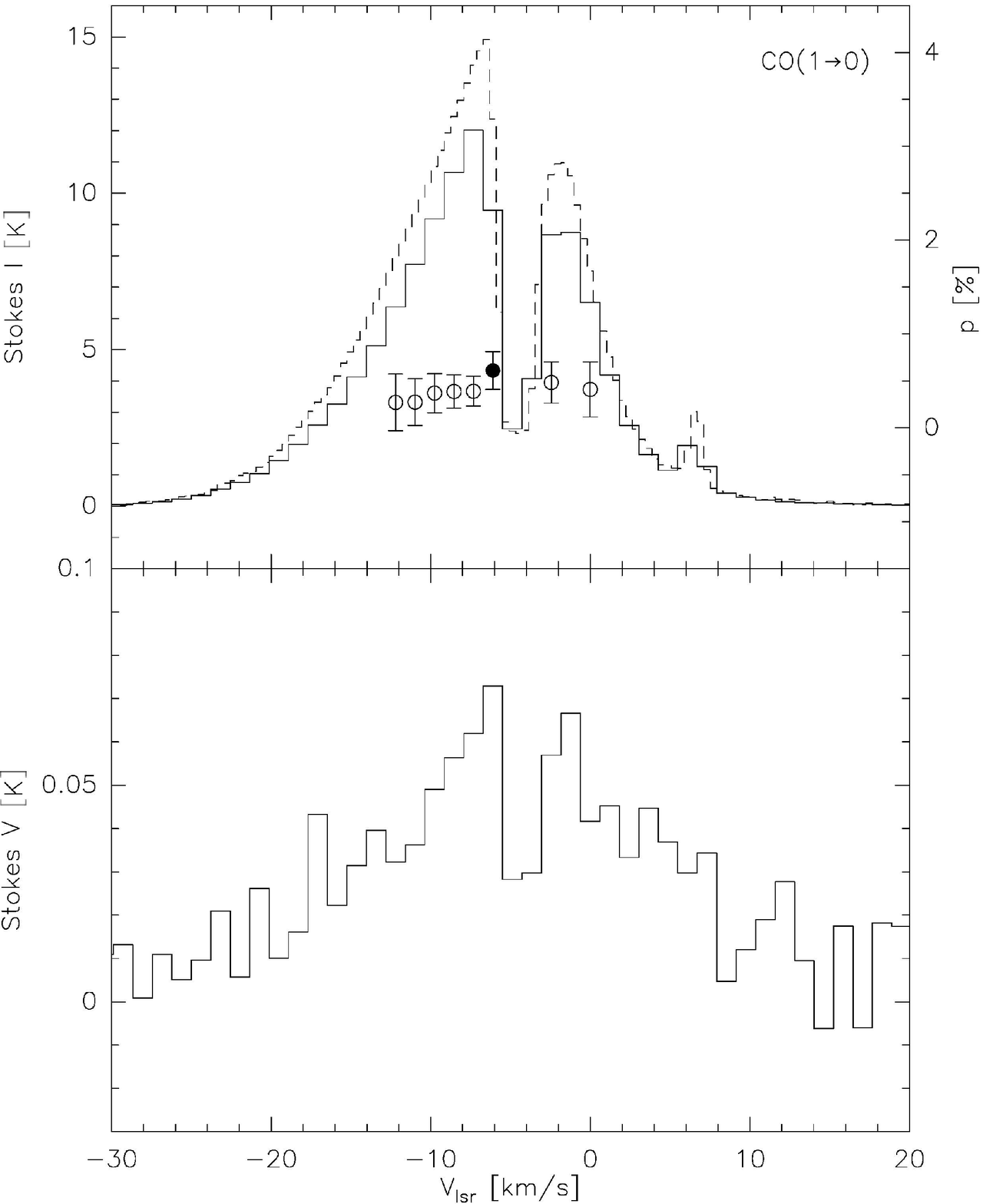}\\ 
\end{tabular}   
\end{center}  
\caption{The circular polarization profiles for
  $^{12}$CO ($J=2\rightarrow1$) (left) and ($J=1\rightarrow0$)
  (right). The Stokes $I$ spectra at offsets ($20^{\prime\prime}$, $60^{\prime\prime}$) from the map
origin ($\alpha\,(J2000)=06^{\mathrm{h}}18^{\mathrm{m}}02\fs7$,
$\delta\,(J2000)=+22^{\circ}39^{'}36^{''}$) overlaid with the fractional circular polarization
  levels are shown in the top panels and the Stokes $V$ spectra
corresponding to the Stokes $I$ profiles are shown in the bottom
panels. These spectra are corrected for instrumental polarization. } 
\label{fig:CO_pC} 
\end{figure*}

In addition to linear polarization, we also detected levels of
circular polarization  in both of the CO transitions in
IC 443-G, which were at times higher than the linear polarization
levels. After the removal of instrumental effects from the 
polarization signals, we obtained intrinsic circular polarization levels of $0.5-1
\%$ across the CO maps. Figure  \ref{fig:CO_pC} shows the Stokes $I$ spectra of
CO ($J=2\rightarrow1$) and ($J=1\rightarrow0$) overlaid with the fractional
circular polarization levels in the top panels at the same position as the data displayed
in Figure \ref{fig:CO_pL}. The Stokes $V$
spectra are plotted in the bottom panels. The instrumental
polarization is removed from all the displayed spectra and the
systematic leakage between the different Stokes parameters (explained
in \S \ref{sec:iram_IP}) are all taken into account. Besides the higher
circular polarization levels across the CO ($J=2\rightarrow1$) line, the
behavior of the Stokes $V$ profile in this transition line is very
different than that of the CO ($J=1\rightarrow0$) line although they
revealed similar linear polarization profiles as seen in Figure \ref{fig:CO_pL}.

The observed circular polarization is not related to the Zeeman effect
since CO is not a Zeeman sensitive molecule. The $^{12}$CO molecule has a Land\'{e}
factor $g^{\mathrm{CO}}_J=-0.269$ \citep{GC84}, which multiplied by the nuclear
magneton gives an approximate Zeeman splitting factor of $0.2$ mHz
$\mu$G$^{-1}$. This splitting is insignificant compared to the
approximately 2 Hz $\mu$G$^{-1}$ factor at 113 and 226 GHz
\citep{BL89} of the CN molecule, which is routinely used for Zeeman
measurements in molecular clouds (\citealt{Crutcher99,Hezareh10}). 

\subsection{Anisotropic Resonance Scattering}

The physics behind the non-Zeeman circular polarization due to
  anisotropic resonant scattering in
 molecular rotational spectral lines was first discussed by
\citet{Houde13}, who performed a
single pointed polarization observation on CO ($J=2\rightarrow1$) in
Orion KL at the Caltech Submillimeter Observatory (CSO) using the
Four-Stokes-Parameter Spectral Line Polarimeter (FSPPol). They detected
non-Zeeman circular polarization levels of up to $2\%$ for this line,
comparable to the linear polarization level that \citet{Girart04} and
\citet{Hezareh10} had measured previously for the same
line at the same offsets in the source.  

The model \citet{Houde13} proposed to explain the presence of
non-Zeeman circular polarization in the spectral lines of molecules
such as CO and SiO is based on the anisotropic resonant scattering of
linear polarization signals incident on these molecules. More
precisely, we may consider a population of such molecules aligned
with the magnetic field in a cloud. The emission from these
molecules can become linearly polarized due to the Goldreich-Kylafis
effect, for example. Upon propagation through the cloud along our line of sight, this polarized emission
becomes incident on similar species of molecules in the foreground.
If the orientation of the magnetic field has changed with respect
to the background, then the radiation is absorbed at a transition 
frequency by the foreground molecules to a virtual state and is
re-emitted via anisotropic resonant scattering with a phase
shift between its scattered amplitudes. The incident and scattered
radiation can be written in terms of $n$-photon states polarized
parallel and perpendicular to the foreground magnetic field projected
in the plane of the sky as \citep{Houde13}
\begin{equation}
\begin{array}{ccc}
\left|\,\psi\right\rangle & = & \alpha \left|\,n_{\Vert}\right\rangle
+ \beta \left|\,n_{\bot}\right\rangle\\
\\
\left|\,\psi'\right\rangle & \simeq & \alpha\,e^{i\phi} \left|\,n_{\Vert}\right\rangle
+ \beta \left|\,n_{\bot}\right\rangle,
\end{array}
\label{eq:psi}
\end{equation}
where $\phi$ is the phase shift between the aforementioned scattered
components and $\theta$ the polarization angle of the incident radiation with respect to
the foreground field ($\alpha=\mathrm{cos}\,(\theta)$ and
$\beta=\mathrm{sin}\,(\theta)$). We can further define $n$-photon 
linear polarization states at $\pm45^{\circ}$ from the field and also
circular polarization states, respectively,  with

\begin{equation}
\begin{array}{ccc}
\left|\,n_{\pm45}\right\rangle = \frac{1}{\sqrt{2}}(\left|\,n_{\Vert}\right\rangle
\pm \left|\,n_{\bot}\right\rangle)\\
\\
\left|\,n_{\pm}\right\rangle = \frac{1}{\sqrt{2}}(\left|\,n_{\Vert}\right\rangle
\pm i\left|\,n_{\bot}\right\rangle),
\end{array}
\label{eq:n}
\end{equation} 
where $\left|\,n_{+}\right\rangle\, (\left|\,n_{-}\right\rangle )$
represents the left (right) handed circular polarization state. Since
the commissioning of XPOL, the polarization measurements at the IRAM
telescope follow the IAU definition (Thum et al. 2008). We adopt this
convention and use the aforementioned states in Equation (\ref{eq:n}) to express
the Stokes parameters as 

\begin{equation}
\begin{array}{l}
I \,\,= \Vert\,\langle\,n_{\Vert}\left|\,\psi'\right\rangle\Vert^2+\Vert\,\langle\,n_{\bot}\left|\,\psi'\right\rangle\Vert^2=\alpha^2+\beta^2\\
\\
Q =
\Vert\,\langle\,n_{\Vert}\left|\,\psi'\right\rangle\Vert^2-\Vert\,\langle\,n_{\bot}\left|\,\psi'\right\rangle\Vert^2=\alpha^2-\beta^2\\
\\
U =
\Vert\,\langle\,n_{+45}\left|\,\psi'\right\rangle\Vert^2-\Vert\,\langle\,n_{-45}\left|\,\psi'\right\rangle\Vert^2=2\alpha\beta\,\mathrm{cos}\,(\phi)\\
\\
V =
\Vert\,\langle\,n_{-}\left|\,\psi'\right\rangle\Vert^2-\Vert\,\langle\,n_{+}\left|\,\psi'\right\rangle\Vert^2=2\alpha\beta\,\mathrm{sin}\,(\phi).\\
\end{array}
\label{eq:stokes}
\end{equation} 

The above equations show that it is the
Stokes $U$ in this specific reference frame that converts to Stokes
$V$ to produce circular polarization,
while Stokes $Q$ and the total amount of polarization remain
unaltered. However, in the absence of anisotropic resonant scattering and
therefore any phase shift $\phi$ between the amplitudes of the
linearly polarized radiation, Stokes $V$ will be zero and the linear 
polarization state of the propagating emission remains unchanged.   

In order to confirm wether it is a conversion of linear to circular
polarization  that gives rise to the observed levels of circular polarization in our data, we first
needed to obtain a dust polarization map of IC 443-G to compare with
the CO polarization maps. We expect the dust polarization vectors to be
perpendicular to the magnetic field lines, while those of CO have the
90$^{\circ}$ ambiguity relative to the magnetic field direction in the
absence of any transformation in the linear polarization states. Therefore, the CO polarization vectors are
ideally expected to be either parallel or perpendicular to the dust
vectors \citep{GK81}. 

\subsection{Linear Polarization Map of Dust Emission}

\begin{figure}
\resizebox{\hsize}{!}{\includegraphics{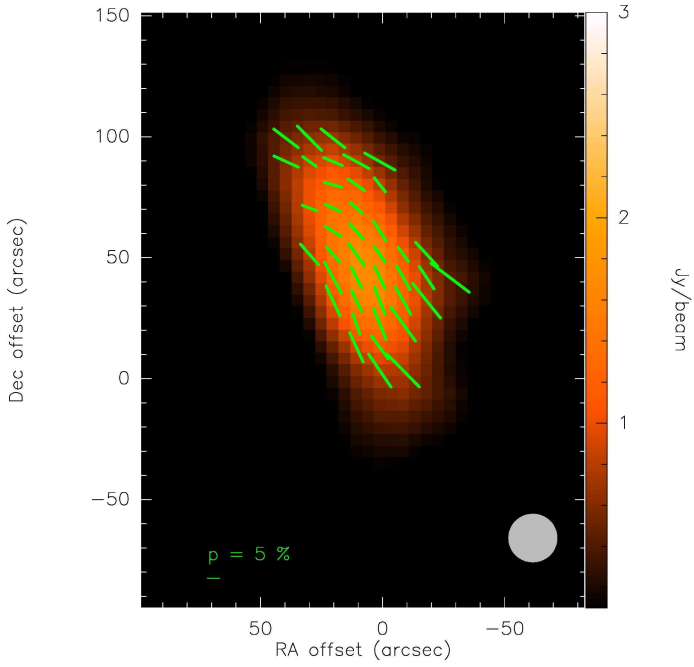}}
\caption{Dust polarization map of IC 443-G at 345 GHz with half-beam
  sampling obtained with PolKa at the
  APEX telescope. The polarization level is up to 10\% and the
  magnetic field is oriented perpendicular to the long axis
  of the source. All the plotted polarization vectors have
  $p\geq3\sigma_p$. The beam size is shown in the lower right corner
  and and the offsets are with respect to the reference coordinates $\alpha\,(J2000)=06^{\mathrm{h}}18^{\mathrm{m}}02\fs7$,
$\delta\,(J2000)=+22^{\circ}39^{'}36^{''}$.}
\label{fig:polka} 
\end{figure}

The dust polarization map of IC 443-G is shown in Figure
\ref{fig:polka}, where the polarization levels are measured to be up
to $10\%$ and the magnetic field seems to be oriented perpendicular to
the long axis of the source. All the plotted and further analyzed
polarization vectors have a fractional linear polarization $p$ of at least
$3 \sigma_p$ and the uncertainty in the polarization angles is less
than $10^{\circ}$.    

\subsection{Estimating the Magnetic Field Orientation in the Foreground}

A pixel-to-pixel comparison between the dust and CO
polarization maps revealed no trend of the corresponding polarization vectors being parallel
or orthogonal. If the inconsistency between the dust and CO 
polarization maps is indeed a result of the transformation from linear
to circular polarization of the latter, then a reverse transformation of these
signals in the appropriate reference frame should recover the original
alignment (parallel or perpendicular) between the dust and CO polarization vectors.  

The first problem is obviously the determination of the orientation
of the foreground magnetic field and distinguishing it from the
background field in this source based on the observations. Our dust polarization map portrays the field orientation
associated with IC 443-G. The CO emission in our maps
originates from the shocked gas, also within the clump but it is not
clear where in the cloud the magnetic
field changes its orientation. 

Given the lack of such information, we performed an
experiment to determine whether there is a unique direction for the magnetic field
that aligns with the CO molecules in the foreground and gives rise to
the measured phase shift of the orthogonal components of the detected CO emission. 
For this purpose, at every pixel of a linear polarization map of CO,
we rotated the reference frame of the Stokes $Q$ and $U$, i.e. the 
equatorial reference frame, from $0^{\circ}$ to $180^{\circ}$  by $5^{\circ}$ steps and calculated
these Stokes parameters, $Q_{\theta}$ and $U_{\theta}$, in the new
frame rotated by angle $\theta$. The corresponding Stokes
$V$ for each pixel was then added to the Stokes $U_{\theta}$, conserving the signs of
both parameters via
\begin{equation}
  U'_{\theta}= U_{\theta}\,\mathrm{cos}\,\phi+V\,\mathrm{sin}\,\phi,
\label{eq:u'}
\end{equation}
 where $\phi=\mathrm{tan^{-1}}(V/U_{\theta})$. 
 The $Q_{\theta}$ and $U'_{\theta}$ parameters where then rotated back to the
 sky coordinates and used to obtain new polarization levels and
 angles. 

\begin{figure*}
\begin{center}
\includegraphics[trim = 0mm 7mm 0mm 10mm, clip,
angle=0,scale=0.5]{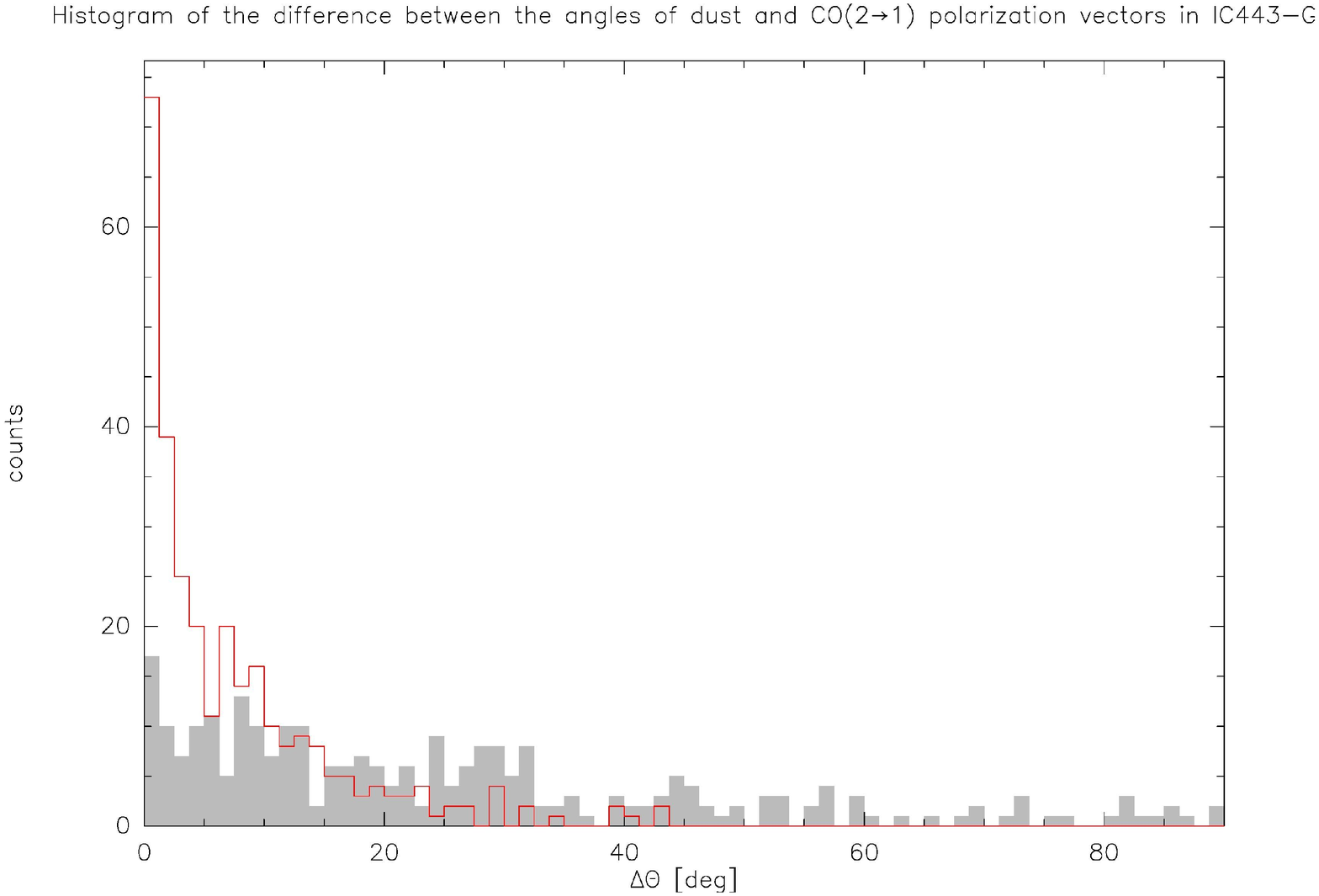}\\ 
\includegraphics[trim = 0mm 0mm 0mm 10mm, clip,
angle=0,scale=0.5]{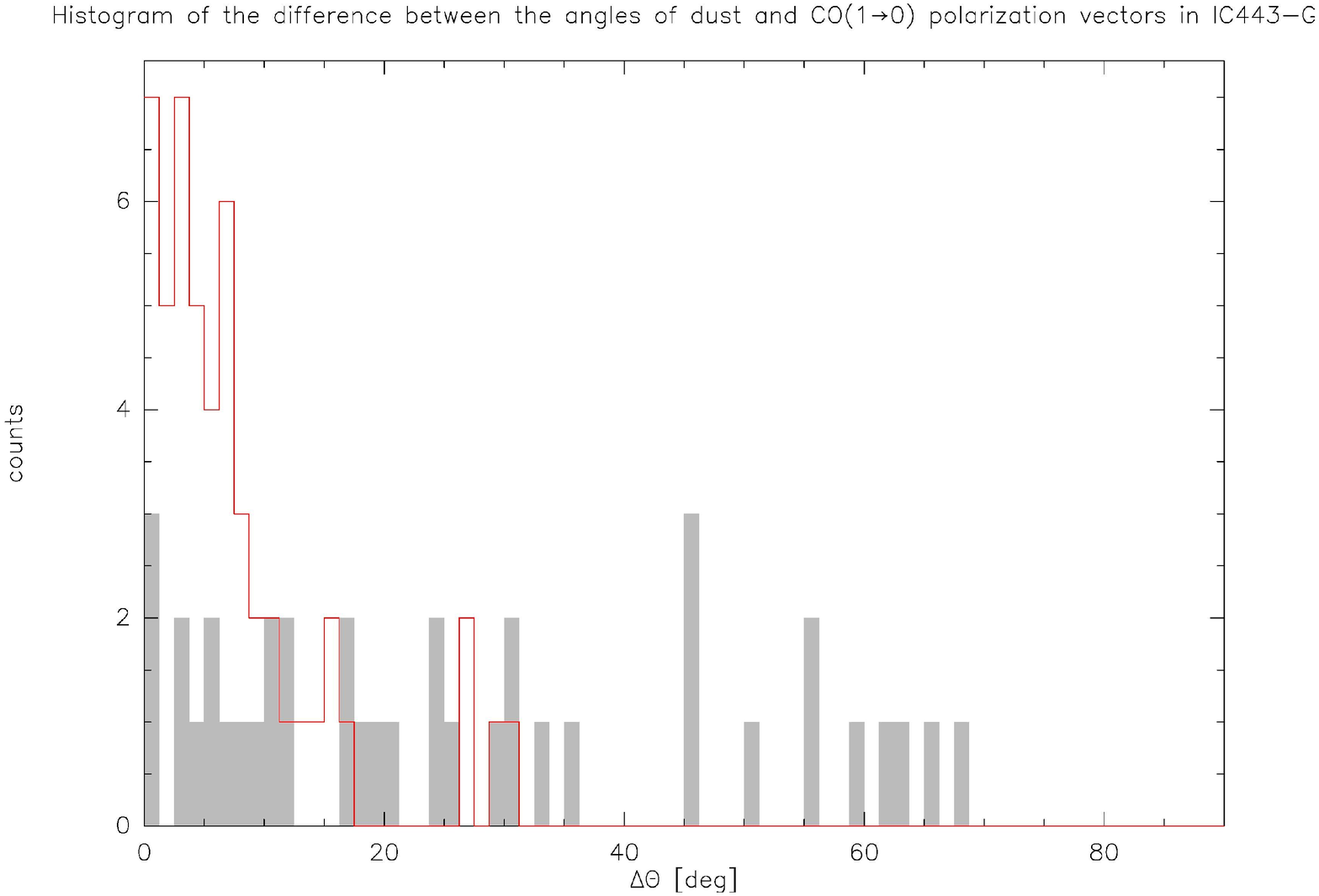}

\end{center}
\caption{(Top): Histograms showing the difference between the polarization
  angles of dust and CO ($J=2\rightarrow1$) maps. The grey histogram is
  a result of comparing the polarization map in Figure \ref{fig:CO21}
  (left) with that of Figure \ref{fig:polka}. In the red histogram the
  Stokes $V$-corrected CO ($J=2\rightarrow1$) polarization map of Figure
  \ref{fig:RS21} is compared to the dust polarization map. The clear
  alignment of the polarization vectors of the two different tracers
  in the red histogram is strong evidence for a conversion of linear
  to circular polarization. (Bottom): Similar histograms for the
  comparison of the polarization vectors of dust and
  CO ($J=1\rightarrow0$) maps.} 
\label{fig:histo} 
\end{figure*}

After this step we produced a histogram of the distribution of the
difference between the angles of the dust and corrected CO
polarization vectors with the aim of finding a peak at either
$0^{\circ}$ or $90^{\circ}$. The top plot in Figure \ref{fig:histo}
shows this histogram obtained with a magnetic field orientation in the
east-west direction on the sky, which yields the
best agreement between the polarization angles of dust and the
blue-shifted wings of CO ($J=2\rightarrow1$) line profiles. For this
field orientation, $73\%$ of the corresponding polarization vectors deviate by
less than $10^{\circ}$, of which $53\%$ differ by less than
$5^{\circ}$. A similar histogram for the same field orientation is shown in the
bottom plot of Figure \ref{fig:histo} on the comparison between polarization angles of dust
and the CO ($J=1\rightarrow0$) transition. Due to the lower resolution
of the latter map, fewer polarization vectors were
available for the statistics, although the histogram shows $76\%$ of
the corresponding polarization vectors deviating by less than
$10^{\circ}$, of which $47\%$ differ by less than $5^{\circ}$. The histograms for the red-shifted
wings of both transitions of CO display a similar trend, showing that
this flux conversion spans a wide range of velocities and that the
polarization 
characteristics of the two CO transitions are the same in this
source. This is seen by comparing the corresponding maps 
for the blue- and red-shifted wings in Figures \ref{fig:CO21} and \ref{fig:CO10},
 which display similar polarization angles for the two cases.

\begin{figure*}
\begin{center}  
\begin{tabular}{cc}  
\includegraphics[trim = 0mm 0mm 6mm 6.8mm, clip, angle=0,scale=0.4]{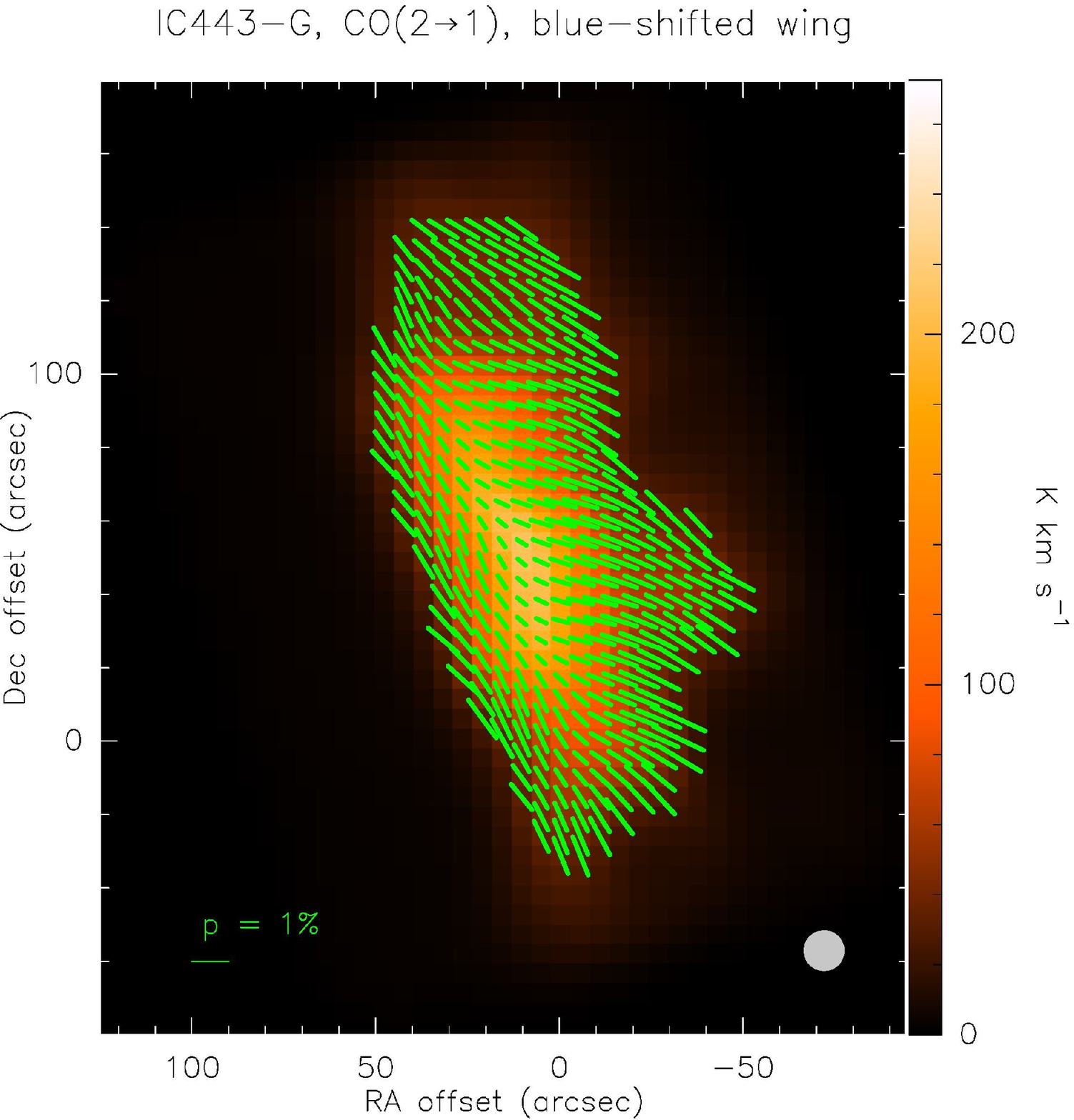}
& \includegraphics[trim = 6.4mm 0mm 0mm 6.9mm, clip, angle=0,scale=0.4]{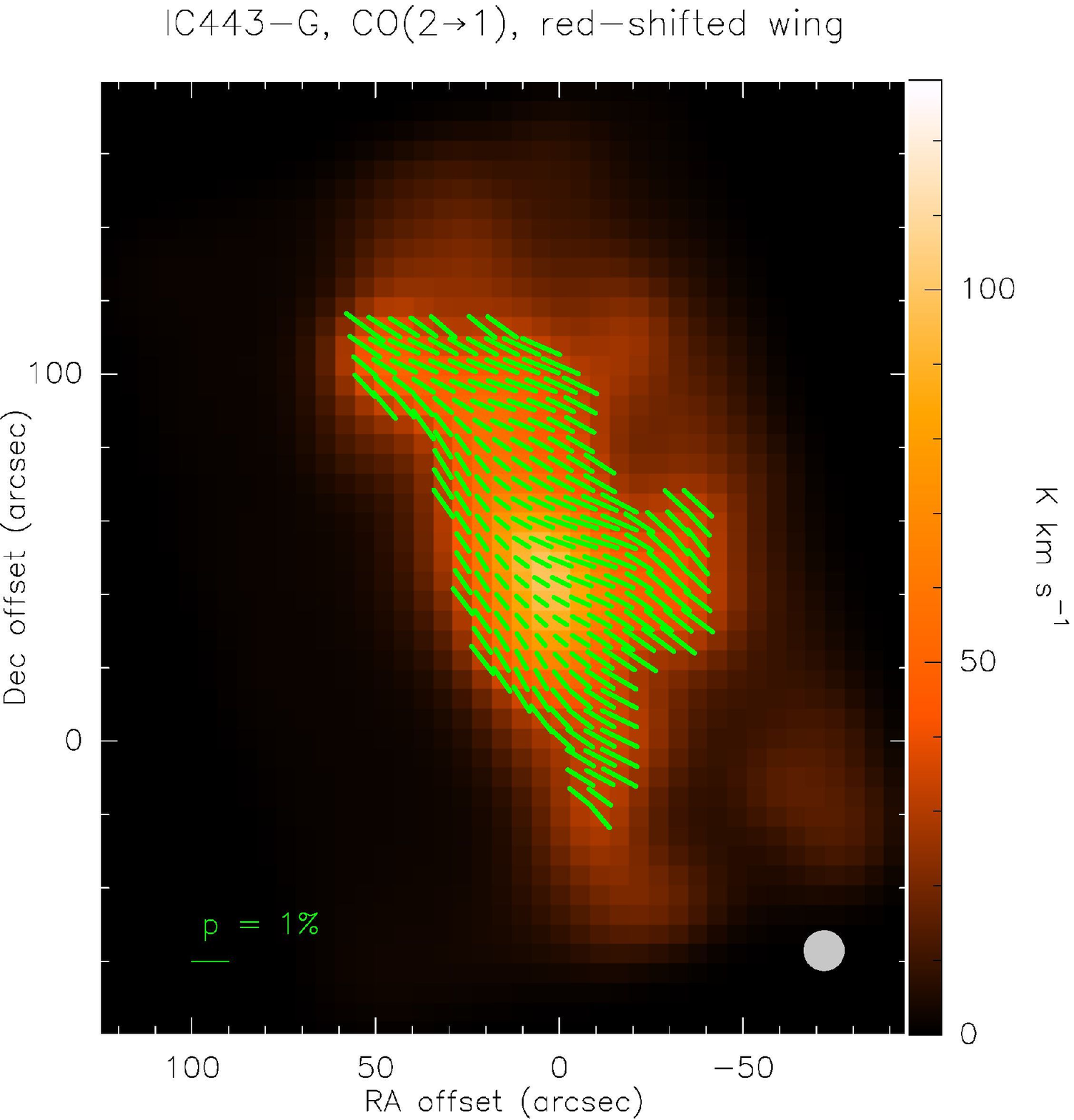} \\ 
\end{tabular}   
\end{center}  
\caption{The polarization maps of CO ($J=2\rightarrow1$) spectral lines
in the blue-shifted (left) and red-shifted (right) wings after the
conversion of the circular polarization levels into linear and re-calculation
of the polarization vectors. Similar to Figures \ref{fig:CO21} and
\ref{fig:CO10}, only polarization vectors with $p\geq3\sigma_p$
are plotted. The beam size for each map is shown in
the lower right corners and the offsets are with respect to the reference coordinates $\alpha\,(J2000)=06^{\mathrm{h}}18^{\mathrm{m}}02\fs7$,
$\delta\,(J2000)=+22^{\circ}39^{'}36^{''}$.}  
\label{fig:RS21} 
\end{figure*}

\begin{figure*}
\begin{center}  
\begin{tabular}{cc}  
\includegraphics[trim = 0mm 0mm 6mm 6.8mm, clip, angle=0, scale=0.4]{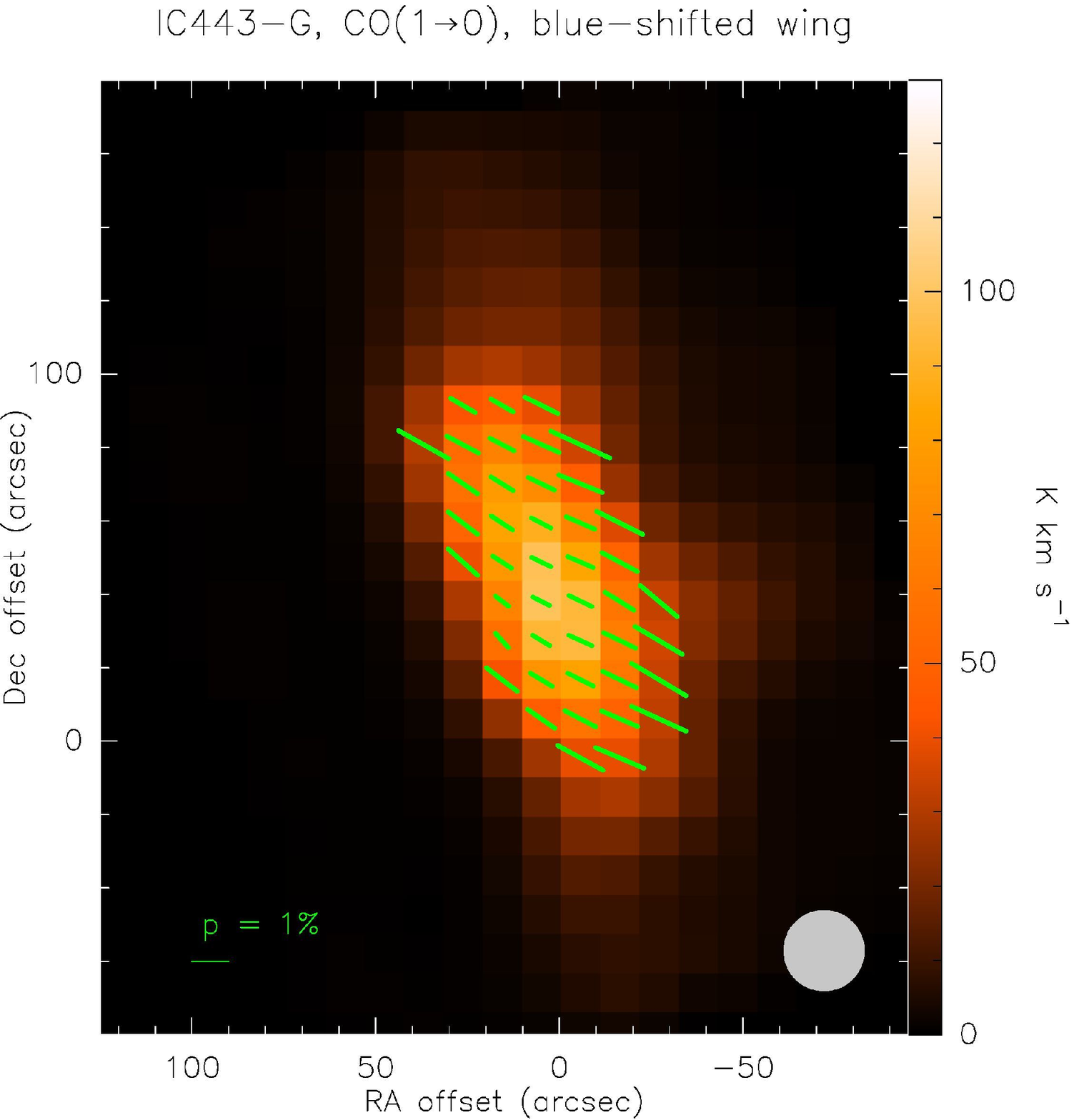}
& \includegraphics[trim = 6.4mm 0mm 0mm 6.9mm, clip, angle=0, scale=0.4]{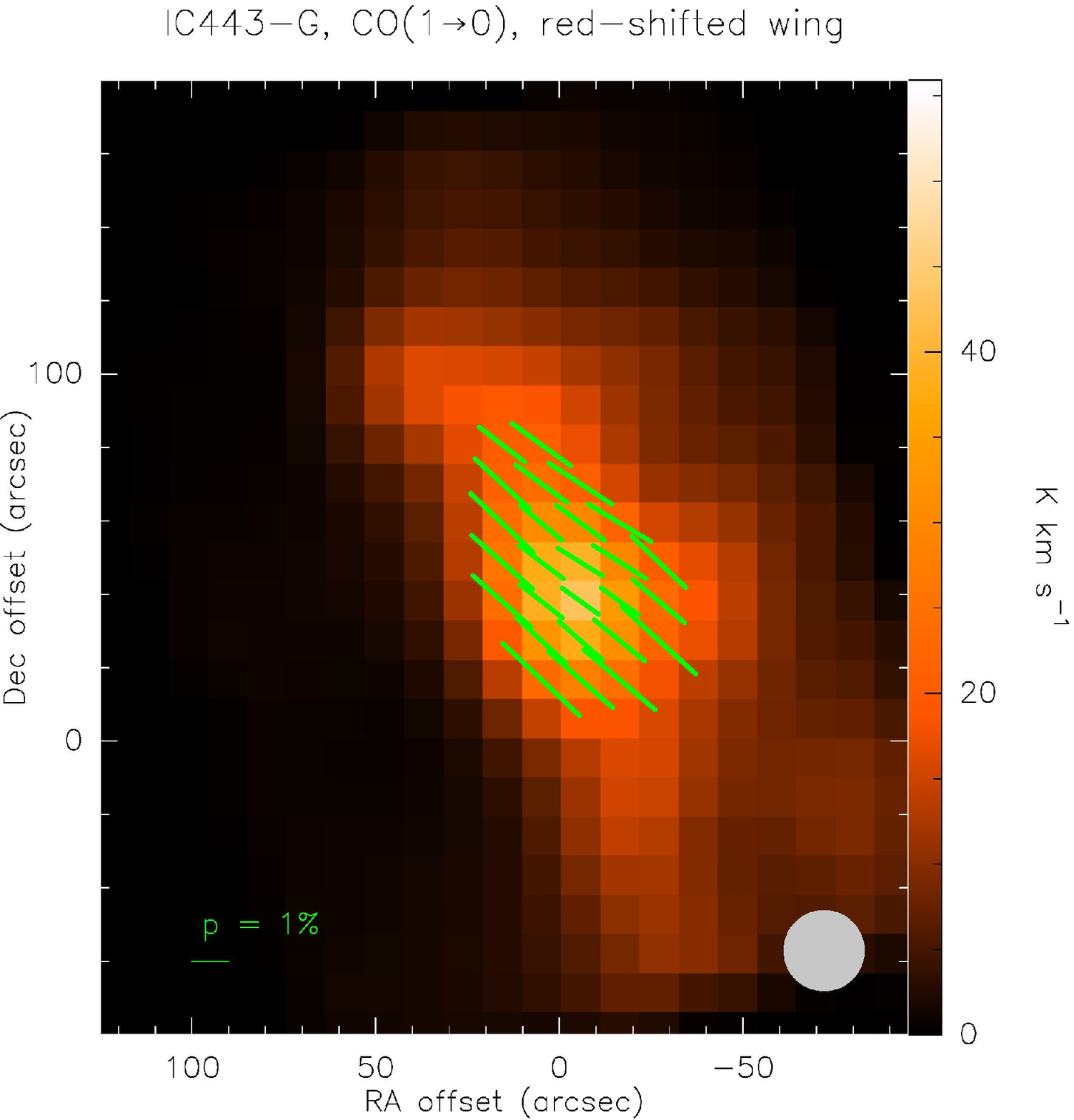} \\ 
\end{tabular}   
\end{center}  
\caption{Same as Figure \ref{fig:RS21} but for the CO($J=1\rightarrow0$) transition.} 
\label{fig:RS10} 
\end{figure*}

With the determined orientation of the foreground field (in the plane
of the sky) we re-plotted the polarization maps for the blue- and red-shifted
wings for both CO transitions to see what linear polarization measurements we would have
obtained in the absence of polarization conversion. Figures
\ref{fig:RS21} and \ref{fig:RS10} show such maps for the
($J=2\rightarrow1$) and ($J=1\rightarrow0$) transitions of CO,
respectively. We note that not only the polarization patterns in the
blue- and red-shifted wings are consistent, they are also in excellent
agreement with the dust polarization map.

\section{Discussion}

This work follows the detection and analysis by
\citet{Houde13} of non-Zeeman circular
polarization in molecular spectral lines. With the aim of studying the magnetic field structure
around the interface of an SNR with
an ambient molecular cloud, we produced polarization maps in both dust
and CO emission in the most massive shocked clump
(G)  towards the western edge of the SNR IC 443 shell. We obtained both
linear and circular polarization levels in the two transitions of CO
and analyzed  our data in several steps to confirm whether our
observational results are consistent with a conversion of linear to circular
polarization and the anisotropic resonant scattering model explained by \citet{Houde13}. 

A very important step in the reduction of polarization data is the
proper removal of instrumental effects from the
observations. Polarization measurements on OTF maps of extended
sources are vulnerable to significant spurious contribution from the polarized
sidelobes of the telescope beam. The amount of this contribution is
variable across the map, and is particularly remarkable towards the
outer edges. We used a robust method of convolving our target map with Stokes maps of
planet Mercury in order to account for and remove these spurious
polarization signals. The intrinsic linear polarization levels of $\simeq0.5-0.8\%$ and
circular polarization levels of up to $\simeq1\%$ that we report in CO
spectral lines are therefore credible results that can be further used
for more in-depth studies of the interaction between the magnetic
field and cloud interface in this SNR. We note that a clear
realization of this polarization conversion in the CO lines would not have been
possible without this correction method for instrumental polarization.

After calibration and removal of instrumental polarization from the
data, we found a unique direction for the magnetic field in the
foreground gas that would provide the best alignment between dust and CO linear
polarization maps after a reverse transformation of circularly polarized
power into linear in the CO data. We have no further information about
the foreground cloud, but we expect that it is optically thin. That is, an incident background
photon cannot be absorbed by the foreground cloud but can undergo
several resonant scattering events before escaping the cloud, whose
thickness should therefore not (approximately) exceed the effective mean
free path set by both the resonant scattering and absorption
processes.

\begin{figure*}
\begin{center}  
\begin{tabular}{cc}  
\includegraphics[trim = 0mm 5.3mm 0mm 0mm, clip, angle=0, scale=0.4]{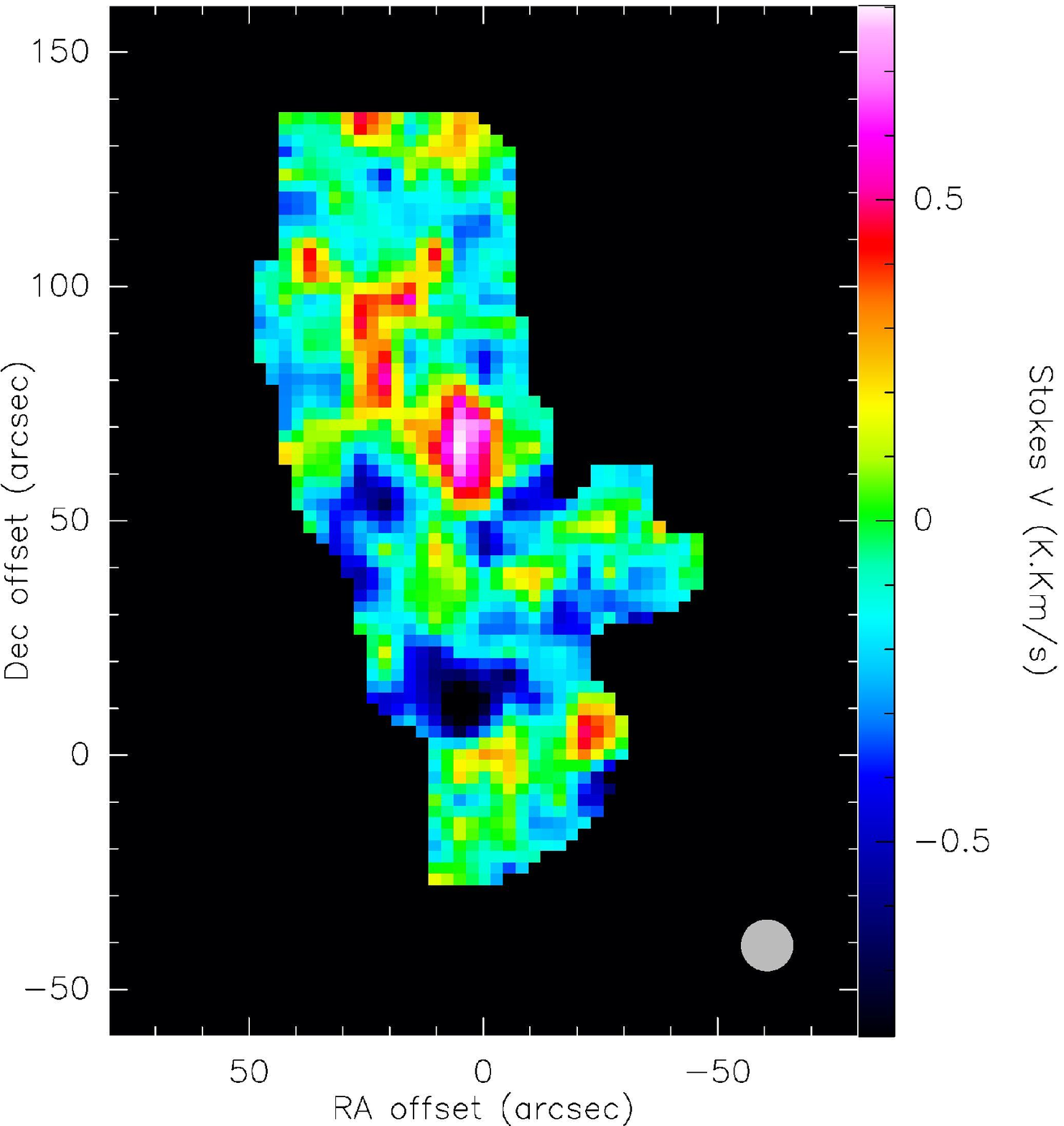}
& \includegraphics[trim = 5.3mm 5.3mm 0mm 0mm, clip, angle=0,
scale=0.4]{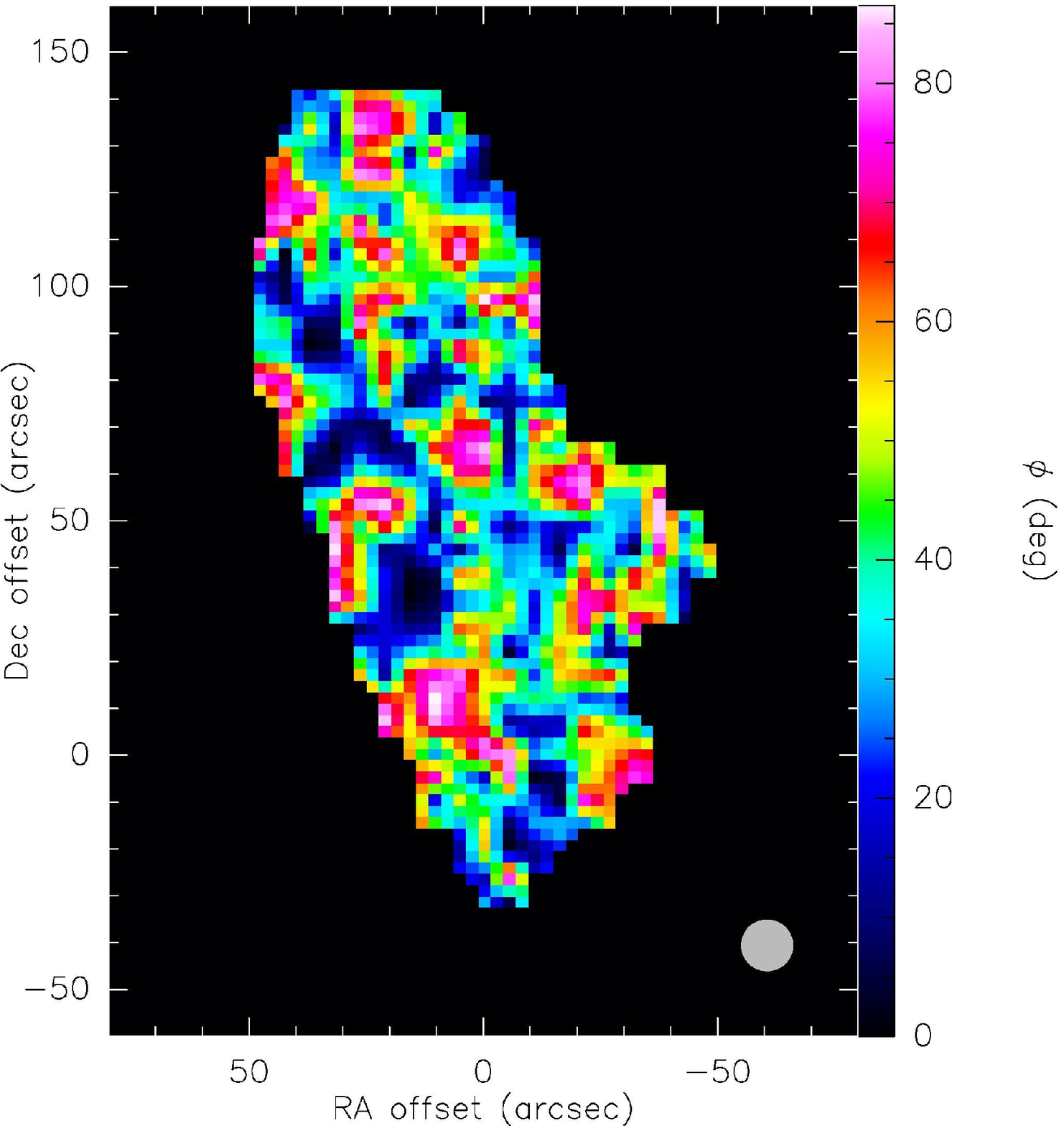} \\ 
\includegraphics[trim = 0mm 0mm 0mm 0mm, clip, angle=0, scale=0.4]{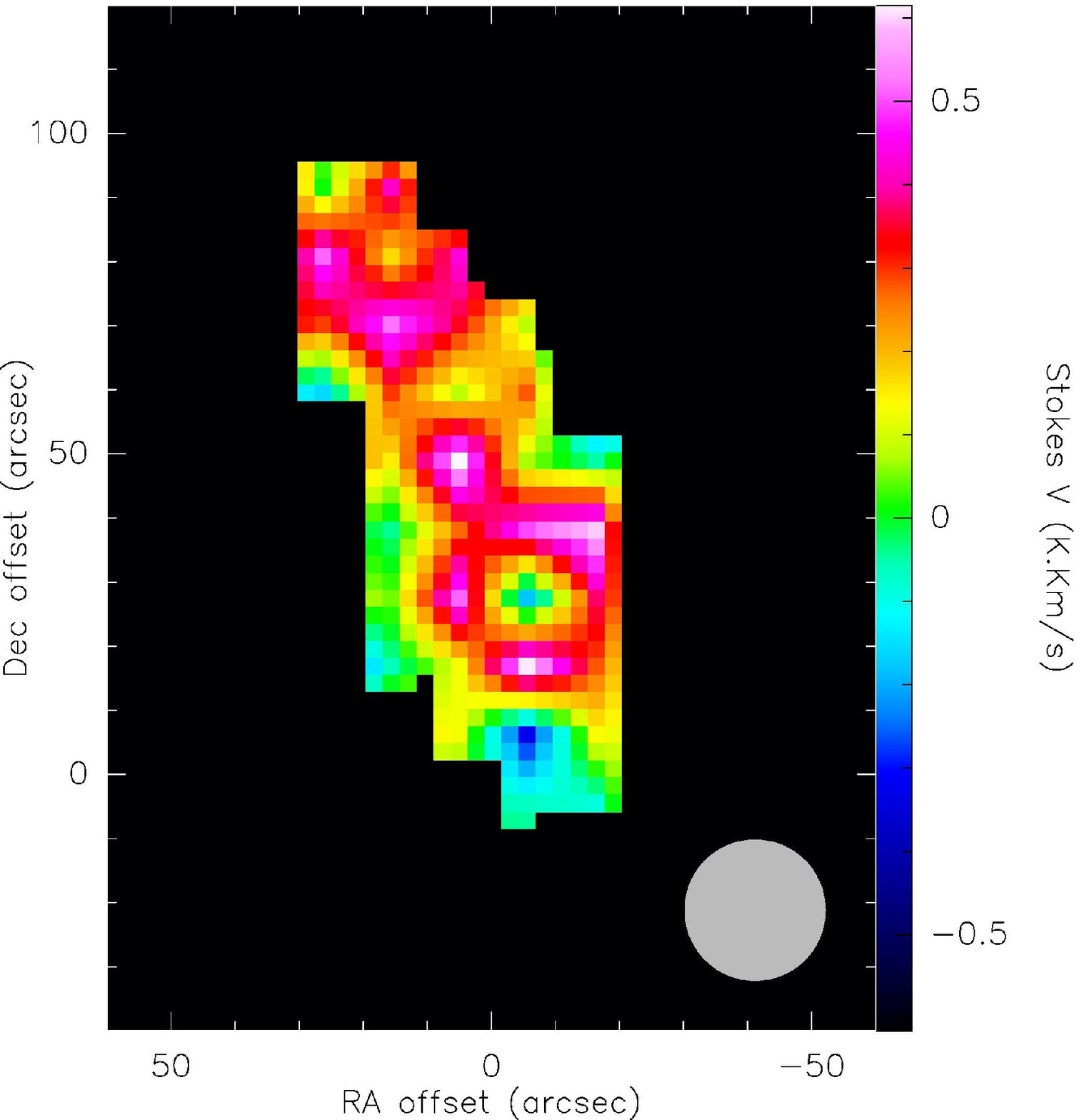}
& \includegraphics[trim = 5.3mm 0mm 0mm 0mm, clip, angle=0, scale=0.4]{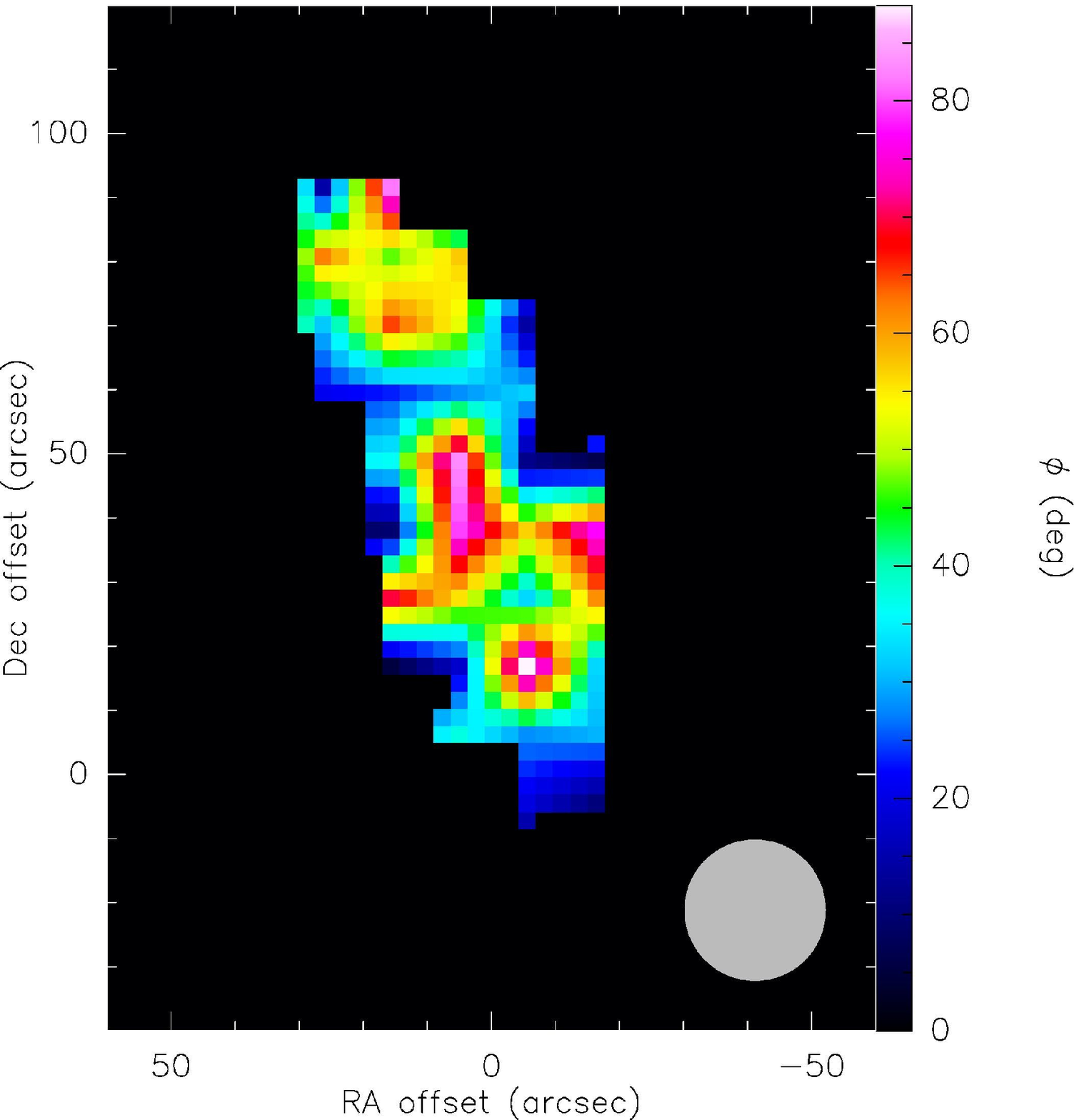} \\ 
\end{tabular}   
\end{center}  
\caption{(Top): The map of Stokes $V$ integrated over the
  blue-shifted wings of the CO ($J=2\rightarrow1$) emission in IC 443-G (left) and the
  corresponding $0^{\circ}$ to $90^{\circ}$ distribution of the
  phase-shift $\phi$ between the orthogonally polarized radiation components of
  CO ($J=2\rightarrow1$) averaged over the same velocity ranges as Stokes
  $V$ (right). The bottom panels show the same as above but
  for CO ($J=1\rightarrow0$). The beam sizes are shown in the lower
  right corners and the offsets are with respect to the reference coordinates $\alpha\,(J2000)=06^{\mathrm{h}}18^{\mathrm{m}}02\fs7$,
$\delta\,(J2000)=+22^{\circ}39^{'}36^{''}$.}
\label{fig:phi_v} 
\end{figure*}

The resulting maps for both the blue- and
red-shifted line wings are very similar for the two transitions,
although their polarization transformation patterns, i.e., the
distribution of the phase shift $\phi$ across the source are very
different. Figure \ref{fig:phi_v} shows the Stokes $V$ maps integrated over the
blue-shifted line wings for the two CO transitions and the
corresponding maps of the phase shift $\phi$ calculated from Equation
(\ref{eq:IQUV}) using the Stokes $U$ and $V$ values integrated across
the blue-shifted line wings. Although the CO ($J=1\rightarrow0$) maps have a lower resolution
compared to the higher transition, it seems that the linear to
circular polarization transformation is more extended in the
former. Moreover, there is no obvious correlation between the
corresponding offsets in the two sets of maps affected by this 
polarization transformation. We therefore believe that we have
obtained a correct direction for the foreground magnetic field, otherwise we
would not have obtained this degree of consistency between the two
sets of CO polarization maps with the dust polarization map (Figure \ref{fig:histo}) after the flux correction.

\citet{Houde13} showed that the magnitude of $\phi$
is proportional to several physical parameters including the
excitation temperature and number density of the observed species and
also the square of the strength of plane-of-sky component of the magnetic
field such that
\begin{equation}
\phi\,(\omega)\simeq 
\omega^2_Z\mathrm{sin}^2(\iota)\,\frac{3\pi\,c^2 l^4_{\mathrm{mp}} A_{\mathrm{ba}}
  n_{\mathrm{CO}}g_1e^{-E_1/kT_{ex}}}{4\hbar\omega^3_0\omega^2Q_{\mathrm{CO}}(T_{\mathrm{ex}})}
\sqrt{u(\omega)u'(\omega)}\,I(\omega),
\label{eq:phi}
\end{equation}
where $\omega_Z$ is the Zeeman splitting, $\iota$ the
magnetic field inclination angle relative to the line of sight,
$l_{\mathrm{mp}}$ the size of the region of interaction,
$n_{\mathrm{CO}}$ the density of CO molecules,
$A_{\mathrm{ba}}$ the spontaneous emission coefficient, $\omega$ and $\omega_0$ the
frequencies of incident photon and the rotational transition,
respectively, $u(\omega)(u'(\omega))$ the incident (scattered) linear
polarization energy density (obtained observationally from the antenna
temperature), and $I(\omega)$ is an integral
over the linear polarization profile for a given CO spectral
line. In order to perform a more in-depth
analysis on the physics of this polarization transformation and its
relationship with the ambient magnetic field, we need to calculate
the temperature and density profiles in this source. This analysis will be the
focus of a follow-up paper. 

Finally, we point out the linear polarization measurements on
 an OH maser line detected at 1.7 GHz by
 \citet{Claussen97} in IC 443-G with the VLA.
They found linear polarization in one out of six observed maser lines, with a polarization level of
4.5 \% and a polarization angle of $21^{\circ} $  (no uncertainty
given). The location of this OH maser line, which \citet{Hoffman03}
revisited, corresponds to offsets
($20^{\prime\prime},60^{\prime\prime}$) from our 
reference coordinates for IC 443-G. At that location we obtained a
polarization angle of $29^{\circ} \pm 2^{\circ} $ from the (corrected)
CO ($J=2\rightarrow1$) polarization map, $32^{\circ}  \pm 3^{\circ} $ from the CO ($J=1\rightarrow0$) map, and
$20^{\circ}  \pm 8^{\circ}$ from our dust polarization
map. Considering the uncertainties in our polarization angles, the
width of the histograms of dust and CO polarization 
angle differences ($5^{\circ}$ for CO ($J=2\rightarrow1$) and $8^{\circ}$ for CO ($J=1\rightarrow0$))
in Figure \ref{fig:histo}, we find our results in good agreement with the
polarization measurement of \citet{Claussen97}. These results also
support their assessment that the observed OH masers indeed arise from
the shocked gas within IC 443-G.

\section{Summary}

We measured linear polarization and also non-Zeeman circular
polarization levels of up to $1\%$ in $^{12}$CO ($J=2\rightarrow1$) and
($J=1\rightarrow0$) spectral line emission observed with the IRAM 30-m
telescope in a shocked molecular clump around the SNR IC 443. We also obtained a dust polarization map of this source
with PolKa at the APEX telescope and compared it to the linear
polarization maps of CO and found no obvious alignment between the corresponding
polarization vectors, suggesting an alteration in the linear
polarization state of the CO emission. We found that the detected circular polarization signal
in CO lines is consistent with a physical model based on anisotropic resonant
scattering introduced by \citet{Houde13}. In this model linearly
polarized CO emission emerging from the background interacts with
similar species of molecules aligned with the ambient magnetic field
in the foreground and scatters at a transition frequency. The relative
phase shift that results between the orthogonally polarized components of this scattered
emission can cause a transformation of linear to circular
polarization. We therefore reinserted the circular polarization signal in the CO
lines across IC 443-G to the corresponding linear polarization
signal to test whether before the linear to circular polarization
conversion the linear polarization vectors of the CO emission were
aligned with those of the dust. Indeed after the flux correction for the two
transitions of the CO spectral lines, the new CO polarization vectors
for both transitions aligned with the dust polarization vectors.

In a future paper we will attempt to establish that the presence of
circular polarization is indeed due to the anisotropic resonant 
scattering model of \citet{Houde13}. If this can be established, then this 
technique would open the possibility to measure the magnetic field in
various astronomical sources using abundant molecules such as CO,
which as explained earlier, is not sensitive to the Zeeman effect. 

\section*{Acknowledgments}

The authors are grateful to C. Thum for very helpful and constructive
comments and discussion. M. Houde's research is funded through the NSERC Discovery Grant, Canada
Research Chair, Canada Foundation for Innovation, and Western's
Academic Development Fund programs. A. Gusdorf acknowledges support by
the grant ANR-09-BLAN-0231-01 from the french Agence Nationale de la
Recherche as part of the SCHISM project.

\clearpage

\newpage

\end{document}